\documentclass[prd,aps,preprint,amsmath,amssymb,]{revtex4-2}
\usepackage{natbib}
\usepackage{booktabs}
\usepackage{dcolumn}
\newcolumntype{d}[1]{D{.}{.}{#1}}
\usepackage{multirow}
\usepackage{amsmath}
\usepackage{amssymb}
\usepackage{graphicx,verbatim}
\usepackage{xcolor}
\usepackage{placeins}
\usepackage{tikz}
\usepackage{tikz-3dplot}
\usetikzlibrary{arrows.meta,calc}
\usepackage[colorlinks=true,linkcolor=red,urlcolor=blue,citecolor=blue]{hyperref}

\graphicspath{{figures/}}
\def\rem#1{{\bf\textcolor{red}{($\diamondsuit$ #1 $\diamondsuit$)}}}

\begin{document}

\title{Construction of Sensitivity Curves for Dynamic LISA and Taiji}

\author{Hong-Yu Shi}
\affiliation{School of Astronomy and Space Science, University of Chinese Academy of Sciences (UCAS), Beijing 100049, China}

\author{Yong Tang}
\affiliation{School of Astronomy and Space Science, University of Chinese Academy of Sciences (UCAS), Beijing 100049, China}
\affiliation{School of Fundamental Physics and Mathematical Sciences, Hangzhou Institute for Advanced Study, UCAS, Hangzhou 310024, China}
\affiliation{International Centre for Theoretical Physics Asia-Pacific, UCAS, Beijing 100190, China}

\begin{abstract}
Space-based gravitational-wave (GW) laser interferometers, including LISA and Taiji, are designed to observe gravitational waves in the millihertz band and are expected to open up a frequency range that is otherwise inaccessible. The sensitivity and response of these instruments are central to their scientific goals, mission design and parameter estimation capabilities. However, they are commonly modeled as static, equilateral triangular constellations, an approximation that neglects both orbital motion and directional dependence. In this work, we systematically examine the direction-dependent response and sensitivity of dynamic LISA-like detectors over an entire year of heliocentric orbit. Based on an analytical, time-dependent heliocentric orbital model and an adiabatic unequal-arm interferometer configuration, we construct direction-dependent sensitivity curves in the Michelson interferometric channel for dynamic LISA and Taiji. We obtain analytic expressions for the angular-dependent sensitivity and demonstrate the emergence of a quadrant-like pattern in sky maps at low frequencies. We show that, relative to the static approximation, the low-frequency sensitivity varies by roughly $20\%$, which in turn produces about a $70\%$ variation in the directional dependence of the number of detectable GW sources, with even larger discrepancies at higher frequencies. Therefore, for accurate predictions of the total GW source counts and reliable parameter inference for binary systems, it is necessary to employ fully dynamic, direction-dependent sensitivity curves.

\newpage
\end{abstract}

\keywords{Gravitational wave; Laser interferometer; Time-delay interferometer}

\maketitle

\section{Introduction}\label{sec:intro}

Space-based gravitational-wave (GW) detectors, such as LISA, Taiji and Tianqin~\cite{LISA, HuWu2017Taiji, TianQin}, are designed to detect gravitational waves (GWs) in the millihertz (mHz) frequency band~\cite{Cutler1998AngularResolution, CornishRubbo2003}.  This frequency band contains a large number of long-lived sources, including massive black-hole binaries, stellar-origin black-hole binaries, extreme mass-ratio inspirals (EMRIs), Galactic compact binaries, and stochastic GW background.  Within this frequency range, the orbital motion of the detector makes the instrumental response intrinsically time dependent.  For a GW source at a fixed sky position, the detector signal is modulated by the orbital motion, by the changing orientation of the triangular constellation, and by the time evolution of the arm lengths between the spacecraft~\cite{Cutler1998AngularResolution, CornishRubbo2003}.

The heliocentric constellation, LISA and Taiji, consists of three spacecraft positioned at the vertices of a nearly equilateral triangle, orbiting the Sun in an Earth-trailing configuration. The constellation also performs a cartwheeling motion, so the detector plane and arm directions vary with respect to a fixed source in the sky~\cite{Cutler1998AngularResolution, CornishRubbo2003}.  
%China's Taiji detector adopts a similar heliocentric triangular configuration, with mission parameters optimized for the millihertz gravitational-wave band~\cite{HuWu2017Taiji}.
The response and sensitivity of LISA-like detectors have been widely analyzed for a static configuration and in an angular-averaged sense~\cite{Tinto:1999yr, Petiteau2008LISACode, Larson2000Sensitivity, CornishRubbo2003, Robson:2018ifk, Smith:2019wny, Zhang:2019oet, Liang:2019pry, Zhang:2020khm, Lu:2019log, Babak:2021mhe, Wang:2021jsv, Wang:2022sti, Du2026TaijiPipeline}. These sensitivity curves are typically obtained by combining instrumental noise models with sky and polarization-averaged response functions.  Such curves provide convenient references for mission design, population studies, and approximate signal-to-noise-ratio (SNR) estimates.   
%The annual periodic motion of the detector induces amplitude and phase modulations in measured signals.  
%Time-Delay Interferometry (TDI) provides the core framework for suppressing laser frequency noise in unequal-arm space interferometers~\cite{TintoDhurandhar2005TDI}, and TDI-based simulators and measurement models form the basis of many modern LISA and Taiji sensitivity calculations~\cite{Petiteau2008LISACode, BayleHartwig2023LISAInstrument, Du2026TaijiPipeline}.
%In many applications, detector performance is commonly simplified into an sky-averaged, static sensitivity curve~\cite{Larson2000Sensitivity, Babak:2021mhe}.  
While this static treatment is suitable for characterizing the overall capability of a mission, for a source at a specific sky location the projection between the source direction and the detector arms changes continuously over the observation time. This gives rise to modulation effects both in the response and in the effective sensitivity~\cite{Adams:2010vc, Wang:2020fwa, Wang:2020pkk}.  Consequently, relying solely on the sky-averaged curves may mask direction-dependent variations, and lead to inaccurate estimations for the total number of GW sources and parameter inference of binary systems.
%when a localized source or a finite observation window is considered.

In this work, we systematically investigate the direction-dependent response and sensitivity curves of dynamic LISA-like detectors over a full-year orbiting period. Using a time-dependent heliocentric orbital model and an adiabatic unequal-arm interferometer structure~\cite{TintoDhurandhar2005TDI, Petiteau2008LISACode}, we construct direction-dependent sensitivity curves in Michelson interferometric channel for dynamic LISA and Taiji. At low frequencies, we are able to analytically derive the angular dependent sensitivity curves and show the quadrant pattern in a sky map. We compare these dynamic sensitivities to the standard static, equal-arm, sky-averaged sensitivities and find that the variation is about $20\%$ in the low-frequency regime, which would lead to $70\%$ variation in the directional dependence of the number of GW sources. And the differences become more pronounced at higher frequencies.
%Then we further examine the dynamic sensitivity to other interferometric channels.

This paper is organized as follows. In Sec.~\ref{sec:theo}, we present the theoretical framework, including the signal response of GWs, the Michelson interferometric channel, and the definition of effective strain sensitivity. In Sec.~\ref{sec:nume}, we describe the numerical implementation, including the orbital model, time averaging, sky sampling, and the fixed-arm all-sky reference calculation. We illustrate with the dynamic sensitivities in Michelson channel, their comparison with the static equal-arm sky-averaged baseline, the low-frequency directional plateaus, and a full-sky sensitivity map at \(3\,\mathrm{mHz}\). We also extend our discussion to other interferometric channels. Finally, in Sec.~\ref{sec:summary} we summarize the main results.

\section{Theoretical Formalism}\label{sec:theo}
This section introduces the signal response and sensitivity framework used
throughout this work. We first define the GW response
of a single inter-spacecraft link and specify the Fourier and delay
conventions. We then construct the Michelson channel \(X\), formulate the conventional static equal-arm sensitivity and extend it to a direction-dependent dynamic sensitivity for an evolving unequal-arm constellation.

\subsection{Signal Response and Interferometric Channels}\label{subsec:single_link}
\begin{figure}[t]
    \centering   
    \includegraphics[width=0.38\linewidth]{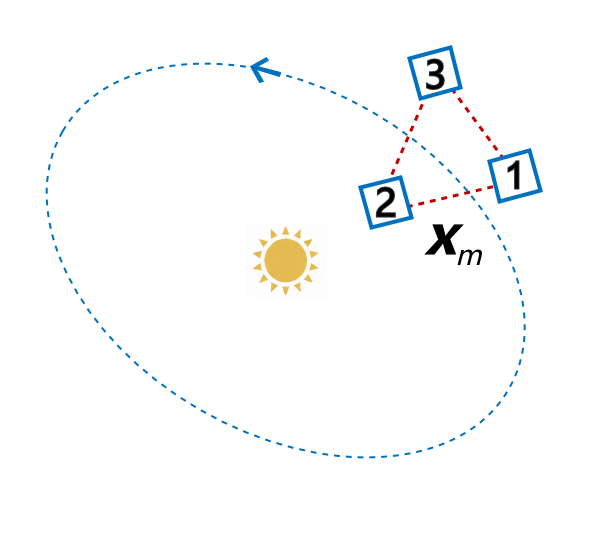}
    \hspace{0.5cm}
    \includegraphics[width=0.55\linewidth]{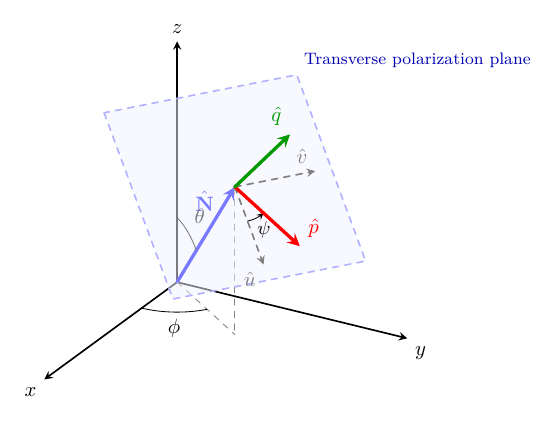}
    \caption{(left) Schematic illustration of the heliocentric triangular constellation. (right) Coordinate and polarization-basis conventions used in the single-link response calculation. The gravitational-wave propagation direction is \(\hat{\Omega}=-\hat{\mathbf N}\), where \((\theta,\phi)\) specify the source direction \(\hat{\mathbf N}\). The vectors \(\hat u\) and \(\hat v\) form a sky-fixed transverse basis, while \(\hat p\) and \(\hat q\) are obtained by a rotation through the polarization angle \(\psi\). }
    \label{fig:coordinates}
\end{figure}
%First, we describe the response of a single inter-spacecraft link to gravitational waves. 
First we introduce the reference frame and conventions in terminology. We denote the unit vector
from the Solar System barycenter toward the source as $\hat{\mathbf{N}}$,
\begin{equation}
    \hat{\mathbf{N}}
    =
    \left(
        \sin\theta\cos\phi,\,
        \sin\theta\sin\phi,\,
        \cos\theta
    \right),
    \label{eq:source_direction}
\end{equation}
and by $\hat{\Omega} = -\hat{\mathbf{N}}$ the propagation direction of the GWs, shown in Fig.~\ref{fig:coordinates}.

In the transverse-traceless gauge, a plane wave propagating along $\hat{\Omega}$ can be written as
\begin{equation}
    h_{ab}(t,\mathbf{x})
    =
    \sum_{A=+,\times}
    h_A
    \left(
        t+\frac{\hat{\Omega}\cdot\mathbf{x}}{c}
    \right)
    e^A_{ab}(\hat{\Omega},\psi),
    \label{eq:plane_gw}
\end{equation}
where \(A=+,\times\) labels the two tensor polarizations,
\(h_A(\tau)\) is the time-domain strain waveform, and
\(\tilde h_A(f)\) denotes its Fourier transform. We use the convention $h_A(\tau)=\int_{-\infty}^{+\infty}\tilde h_A(f)e^{-2\pi i f \tau}\,\mathrm{d}f .$
\begin{align}
    \hat{u}
    &=
    \left(
        \cos\theta\cos\phi,\,
        \cos\theta\sin\phi,\,
        -\sin\theta
    \right),
    \\
    \hat{v}
    &=
    \left(
        -\sin\phi,\,
        \cos\phi,\,
        0
    \right).
\end{align}
These vectors satisfy
\begin{equation}
    \hat{u}\cdot\hat{\Omega}
    =
    \hat{v}\cdot\hat{\Omega}
    =
    0,
    \qquad
    \hat{u}\times\hat{v}
    =
    \hat{\mathbf{N}}
    =
    -\hat{\Omega}.
\end{equation}
The polarization basis $(\hat{p},\hat{q})$ is obtained by rotating
$(\hat{u},\hat{v})$ through the polarization angle $\psi$,
\begin{align}
    \hat{p}
    &=
    \hat{u}\cos\psi
    +
    \hat{v}\sin\psi,
    \\
    \hat{q}
    &=
    -\hat{u}\sin\psi
    +
    \hat{v}\cos\psi.
\end{align}
The corresponding polarization tensors are
\begin{equation}
    e^+_{ab}
    =
    \hat{p}_a\hat{p}_b
    -
    \hat{q}_a\hat{q}_b,
    \qquad
    e^\times_{ab}
    =
    \hat{p}_a\hat{q}_b
    +
    \hat{q}_a\hat{p}_b.
    \label{eq:polarization_tensors}
\end{equation}
\begin{comment}
The angles \(\theta\) and \(\phi\) specify the source direction \(\hat{\mathbf N}\), while the gravitational-wave propagation direction is \(\hat{\Omega}=-\hat{\mathbf N}\). Because \(\hat u\) and \(\hat v\) are orthogonal to \(\hat{\Omega}\), they span the transverse polarization plane. The physical polarization basis \((\hat p,\hat q)\) is obtained by rotating this sky-fixed reference basis through the polarization angle \(\psi\), as illustrated in Fig.~\ref{fig:coordinates}. In the single-link expressions below, the \(+\) and \(\times\) response components are firstly evaluated in the reference basis \((\hat u,\hat v)\), corresponding to \(\psi=0\). A general polarization angle is subsequently introduced through the rotation in Eq.~\eqref{eq:polarization_angle_response}. The polarization-averaged response is independent of the choice of reference transverse basis.    
\end{comment}

Let $\mathbf{x}_i(t)$ be the barycentric position of spacecraft $i$. For a laser link between spacecraft $i$ and $j$, the instantaneous arm length and unit link vector are
\begin{equation}
 L_{ij}(t)=|\mathbf{x}_j(t)-\mathbf{x}_i(t)|,
 \qquad
 \hat{n}_{ij}(t)=\frac{\mathbf{x}_j(t)-\mathbf{x}_i(t)}{L_{ij}(t)} .
 \label{eq:arm_definition}
\end{equation}
In this paper, we use the rigid adiabatic approximation, which means we assume the constellation geometry remains static during a single light-travel time, yet we still account for orbital effects on an annual timescale~\cite{Petiteau2008LISACode}. 

We adopt the link convention that $y_{ij}$ denotes a measurement of fractional frequency change for the laser light received at
spacecraft $i$ from the sending spacecraft $j$. Following the standard one-way Doppler response \cite{EstabrookWahlquist1975,Vallisneri2005} and its rigid-adiabatic frequency-domain implementation \cite{RubboCornishPoujade2004}, for a monochromatic plane wave with frequency $f$, the gravitational-wave contribution to a single-link measurement can be written as
\begin{equation}
 y_{ij}(f,t,\hat{\Omega})=
 \sum_A \xi^A_{ij}(t,\hat{\Omega})
 \left[
 e^{-2\pi i f\,\hat{\Omega}\cdot\mathbf{x}_i(t)/c}
 -
 e^{+2\pi i f L_{ij}(t)/c}
 e^{-2\pi i f\,\hat{\Omega}\cdot\mathbf{x}_j(t)/c}
 \right]\tilde{h}_A(f)e^{-2\pi i f t},
 \label{eq:onelink_phase}
\end{equation}
where the geometric projection factors are
\begin{equation}
 \xi^+_{ij}=-\frac{1}{2}\frac{(\hat{u}\cdot\hat{n}_{ij})^2-(\hat{v}\cdot\hat{n}_{ij})^2}{1-\hat{\Omega}\cdot\hat{n}_{ij}},
 \qquad
 \xi^\times_{ij}=-\frac{(\hat{u}\cdot\hat{n}_{ij})(\hat{v}\cdot\hat{n}_{ij})}{1-\hat{\Omega}\cdot\hat{n}_{ij}} .
 \label{eq:xi_projection}
\end{equation}
%We adopt the Fourier convention $e^{-2\pi i f t}$, for which a time delay of $L_{ij}/c$ corresponds in the frequency domain to a multiplicative factor of $\exp(+2\pi i f L_{ij}/c)$. 
%The convention in Eq.~\eqref{eq:onelink_phase} is employed consistently in both the dynamic response calculation and the fixed-arm all-sky reference.
From Eq.~\eqref{eq:onelink_phase}, it can be seen that, in addition to the phase factor, $e^{-2\pi i f t}$, the time dependence of the single-link response arises from the orbital evolution of the triangular constellation and is encoded in the time-dependent spacecraft positions, instantaneous arm lengths, and link directions. 

%\subsection{Interferometric Channels}\label{subsec:michelson_x}
The above single-link measured quantities are overwhelmed by the laser frequency noise, which motivates the construction of time-delay interferometry channels with virtual equal-arm light paths~\cite{Tinto:1999yr, Tinto:2002de, TintoDhurandhar2005TDI, Vallisneri:2005ji}. For conventional sensitivity curves, one usually considers the Michelson \(X\) channel, whose construction is as follows. We define the two round-trip combinations
\begin{equation}
 \eta_{12}=y_{12}+\mathcal{D}_{12}y_{21},
 \qquad
 \eta_{13}=y_{13}+\mathcal{D}_{13}y_{31} .
 \label{eq:eta_def}
\end{equation}
Here the delay operator is defined as $\mathcal{D}_{ij}y_{mn}(t)=y_{mn}\left(t-{L_{ij}(t)}/{c}\right)$.
The compact unequal-arm Michelson channel is then
\begin{equation}
 X(t)=\left(1-\mathcal{D}_{13}\mathcal{D}_{31}\right)\eta_{12}
 -\left(1-\mathcal{D}_{12}\mathcal{D}_{21}\right)\eta_{13}.
 \label{eq:TDI_X_compact}
\end{equation}
 $Y$ and $Z$ can be obtained by index cycle $(1\rightarrow 2 \rightarrow 3 \rightarrow 1)$. Here we concentrate on the above first-generation interferometric channels, since our interest lies in their sensitivities, which are identical to those of the corresponding second-generation channels~\cite{Tinto:2022zmf}. 

Note that for fully time-dependent arms, the exact delay operators do not
commute. In the frozen-arm adiabatic calculation used here, however, the constellation is held fixed during each instantaneous frequency-domain evaluation, and the delay operators are represented by commuting scalar factors $D_{ij}(f,t)=\exp[+2\pi i fL_{ij}(t)/c]$.

%In the frequency-domain adiabatic implementation, each delay operator is replaced by the scalar factor $D_{ij}=\exp(+2\pi i fL_{ij}/c)$.

\subsection{Effective strain sensitivity}
\label{subsec:sensitivity}

For an interferometric channel $I$, the frequency-domain output is written as
\begin{equation}
\tilde{s}_I(f;t)
=
\tilde{h}_I(f;t,\hat{\Omega})
+
\tilde{n}_I(f;t),
\label{eq:tdi_output}
\end{equation}
where \(\tilde{h}_I\) and \(\tilde{n}_I\) denote the
gravitational-wave and instrumental-noise contributions, respectively. The main instrumental noises are the optical metrology system (OMS) noise and the residual acceleration noise of the test
masses. Here \(t\) labels the slowly evolving orbital configuration of the detector rather than the Fourier-transform variable. At each epoch, the constellation is treated as frozen within the adiabatic approximation.  The signal part can be decomposed as
\begin{equation}
\tilde{h}_I(f;t,\hat{\Omega})
=
\mathcal{F}_{I,+}(f;t,\hat{\Omega})\tilde{h}_+(f)
+
\mathcal{F}_{I,\times}(f;t,\hat{\Omega})\tilde{h}_\times(f).
\label{eq:channel_response_tensor}
\end{equation}
where \(\mathcal{F}_{I,+}\) and \(\mathcal{F}_{I,\times}\) are the complex frequency-domain antenna
response functions of channel \(I\) to the \(+\) and \(\times\)
polarizations.
For a general polarization with angle $\psi$, we have \begin{equation}
\mathcal{F}_I(f;t,\hat{\Omega},\psi)
=
\mathcal{F}_{I,+}(f;t,\hat{\Omega})\cos 2\psi
+
\mathcal{F}_{I,\times}(f;t,\hat{\Omega})\sin 2\psi .
\label{eq:polarization_angle_response}
\end{equation}
Then averaging over $\psi\in[0,\pi)$ gives the unpolarized response function
\begin{equation}
{\mathcal{R}}_I(f;t,\hat{\Omega})
\equiv
\frac{1}{\pi}
\int_0^\pi
\left|
\mathcal{F}_I(f;t,\hat{\Omega},\psi)
\right|^2
\mathrm{d}\psi
=
\frac{1}{2}
\left(
|\mathcal{F}_{I,+}|^2+
|\mathcal{F}_{I,\times}|^2
\right).
\label{eq:polarization_average_discrete}
\end{equation}
This quantity is the half trace of the response in the
two-dimensional tensor-polarization space and is invariant under a
rotation of the polarization basis.

At a fixed orbital time, the effective strain-noise PSD is
\begin{equation}
S_{I}(f;t,\hat{\Omega})
=
\frac{
P_I(f;t)
}{
{\mathcal{R}}_I(f;t,\hat{\Omega})
},
\label{eq:single_channel_sensitivity}
\end{equation}
where $P_I(f;t)$ is the instrumental channel-noise PSD, see the appendix for details. 
In the static equal-arm reference, \(P_I\) is time independent. In the
dynamic unequal-arm calculation, the underlying noises from optical metrology system and acceleration of test masses
are taken to be stationary, while the noise transfer functions may acquire a weak time dependence through the slowly varying
arm lengths.
The connection between \(\tilde h_I(f)\) and \(S_I\) is that \(S_I\) is the
effective strain-noise PSD appearing in the SNR integral. Here \(\rho_I\) denotes the matched-filtering SNR,
\begin{equation}
    \rho_I^2
    \simeq
    4\int
    \frac{
        |\tilde h(f)|^2
    }{
        S_I(f;t,\hat{\Omega})
    }
    \,{\rm d}f ,
    \label{eq:snr_effective_strain}
\end{equation}
up to convention-dependent normalization factors. Because $\rho_I=1$ is used to set the sensitivity curve, as a function of monochromatic wave $|\tilde h_A(f)|$, one may use $\tilde h_A(f) \simeq S_I^{1/2}$ for quick estimation.
%Thus \(S_I^{1/2}\) has the same strain-amplitude units as the source spectrum \(|\tilde h(f)|\) per square-root frequency, and provides the noise level against which the source strain is compared. \rem{}

\begin{comment}
The amplitude spectral density displayed in the figures is
\begin{equation}
S_{I}^{1/2}(f;t,\hat{\Omega})
=
\left[
\frac{
P_I(f;t)
}{
{\mathcal{R}}_I(f;t,\hat{\Omega})
}
\right]^{1/2}.
\label{eq:single_channel_sensitivity_asd}
\end{equation}    
\end{comment}

For the static equal-arm model, we neglect the time dependence of all quantities (except $\exp{(-2\pi i f t)}$) in Eq.~\ref{eq:onelink_phase} and set their values at some reference time $t=t_0$. In such a case the sky-averaged response is defined as
\begin{equation}
\overline{\mathcal{R}}_I^{\mathrm{stat}}(f)
=
\frac{1}{4\pi}
\int
{\mathcal{R}}_I(f;t_0, \hat{\Omega})\,\mathrm{d}\hat{\Omega}.
\label{eq:all_sky_response}
\end{equation}
Its effective strain-noise PSD is
\begin{equation}
S_{I}^{\mathrm{stat}}(f)
=
\frac{P^{stat}_I(f)}{\overline{\mathcal{R}}_I^{\mathrm{stat}}(f)}.
\label{eq:all_sky_sensitivity}
\end{equation}
%Since both quantities are time independent, the ratio-of-averages and SNR-equivalent definitions coincide for the static reference.

For a dynamic constellation orbiting the Sun over the observation time \(T_{\rm obs}\) (we shall take it to be one year for sensitivity curves), we define the time-averaged response and channel-noise PSD as
\begin{equation}
\overline{\mathcal R}^{\rm dyn}_I(f;\hat\Omega)
=
\frac{1}{T_{\rm obs}}
\int_0^{T_{\rm obs}}
\mathcal R_I(f;t,\hat\Omega)\,dt,
\qquad
\overline P^{\rm dyn}_I(f)
=
\frac{1}{T_{\rm obs}}
\int_0^{T_{\rm obs}}
P_I(f;t)\,dt .
\label{eq:dynamic_average_response_noise}
\end{equation}
Here the noise PSD is independent of the source direction, but can acquire a
weak time dependence through the slowly varying unequal-arm noise-transfer
functions.

Then the annual angle-dependent sensitivity is given by
\begin{equation}
S_{I}^{\mathrm{dyn}}(f;\hat{\Omega})
=
\frac{\overline{P}^{\mathrm{dyn}}_I(f)}{
\overline{\mathcal{R}}_I^{\mathrm{dyn}}(f;\hat{\Omega})
}.
\label{eq:dynamic_sensitivity_single}
\end{equation}
This ratio-of-annual-averages definition preserves the usual
noise-to-response form and makes it possible to separate the
unequal-arm noise-transfer correction from the directional response
correction.

\section{Numerical Results}
\label{sec:nume}
\label{sec:results}
\begin{comment}
\rem{,}The calculation is organized around two self-consistent detector models.  The first is the direction-dependent dynamic unequal-a
rm model, in which the spacecraft positions, arm lengths, link directions, responses, and TDI noise transfer functions evolve along a heliocentric orbit.  The second is a fixed equal-arm model used to construct the conventional static all-sky averaged sensitivity.  The two models use the same one-link phase convention, TDI definitions, and underlying OMS and acceleration-noise spectra.  Their effective sensitivities differ, however, through both the response functions and the arm-dependent TDI noise transfer functions.
\end{comment}
In this section, we shall implement our formalism by numerical simulations. After describing the orbits of three spacecraft and the relevant parameters, we numerically calculate the response function and the sensitivity curves for dynamic Taiji and LISA. In the low-frequency regime, we are able to analytically calculate the sensitivity curves for a general direction. Then we also extend the sensitivity curves for other interferometric channels and discuss the implications of these results.

%\subsection{Parameter Setup}\label{subsec:numerical_orbit}
\subsection{Dynamic sensitivities for Michelson Channel}

The spacecraft orbits are modeled using a standard analytic heliocentric triangular configuration with a fiducial arm length $L$ ($L=2.5\times10^9\,\mathrm{m}$ for LISA and $L=3.0\times10^9\,\mathrm{m}$ for Taiji)~\cite{Cutler1998AngularResolution, CornishRubbo2003, Petiteau2008LISACode}. The guiding center follows a heliocentric orbit with radius $R=1\,\mathrm{AU}$ and period $T_{\rm yr}=365\times24\times3600\,\mathrm{s}$. The orbital eccentricity is chosen as $e=\frac{L}{2\sqrt{3}R}$, so that the mean inter-spacecraft separation is approximately $L$. The orbital phase is denoted by $\alpha(t)=\frac{2\pi t}{T_{\rm yr}}+\alpha_0$, where $\alpha_0$ is the initial orbital phase. 

Following the standard analytic heliocentric orbit model expanded to
second order in the orbital eccentricity~\cite{RubboCornishPoujade2004},
the barycentric positions of spacecraft $m=1,2,3$ are written as
$\mathbf{x}_m(t)=\bigl(x_m(t),y_m(t),z_m(t)\bigr)$,
\begin{align}
x_m
={}& R\cos\alpha
+\frac{eR}{2}
\left[
\cos(2\alpha-\beta_m)-3\cos\beta_m
\right] \nonumber \\
&+
\frac{e^2R}{8}
\left[
3\cos(3\alpha-2\beta_m)
-10\cos\alpha
-5\cos(\alpha-2\beta_m)
\right],
\label{eq:orbit_x}
\\
y_m
={}& R\sin\alpha
+\frac{eR}{2}
\left[
\sin(2\alpha-\beta_m)-3\sin\beta_m
\right] \nonumber \\
&+
\frac{e^2R}{8}
\left[
3\sin(3\alpha-2\beta_m)
-10\sin\alpha
+5\sin(\alpha-2\beta_m)
\right],
\label{eq:orbit_y}
\\
z_m
={}&
-\sqrt{3}\,eR\cos(\alpha-\beta_m) \nonumber \\
&+
\sqrt{3}\,e^2R
\left[
\cos^2(\alpha-\beta_m)
+2\sin^2(\alpha-\beta_m)
\right].
\label{eq:orbit_z}
\end{align}
Here the relative phases of the three spacecraft are
$ \beta_1=\frac{2\pi}{3}, 
\beta_2=\frac{4\pi}{3}, 
\beta_3=2\pi . $
The instantaneous arm lengths and link directions are then computed from Eq.~\eqref{eq:arm_definition}. In this manner, the slowly changing arm lengths and link orientations are refreshed throughout the yearly orbit. The numerical orbits with precise solar system dynamics would deviate from the above analytical ones at percent level, which only affect the results at similar level and would not change our main conclusions. 

For every direction in the sky, the response is assessed on a daily basis.  Within each day, the one-link responses are sampled every $\Delta t=0.5\,\mathrm{s}$ and averaged to obtain a daily response curve.  The dynamic response is then obtained by averaging the daily response curves over the full year.
\begin{align} 
\overline{\mathcal{R}}_{I}^{\rm dyn}(f;\hat{\Omega}) &= \frac{1}{N} \sum_{i=0}^{N-1} {\mathcal{R}}_{I}(f;t_i,\hat{\Omega}), \label{eq:annual_response_average_discrete}
\end{align}
Here $N$ is the total number of samples. This procedure retains the slow orbital breathing of the constellation while using the adiabatic approximation over individual light-travel times.

For comparison, we also compute the static sky-averaged response by numerical integration over a midpoint sky grid
\begin{equation}
 \overline{\mathcal{R}}_X^{\rm stat}(f)
 \simeq
 \frac{1}{4\pi}
 \sum_{a,b}{\mathcal{R}}_X(f;\theta_a,\phi_b)\sin\theta_a\,\Delta\theta\,\Delta\phi .
 \label{eq:sky_discrete}
\end{equation}
Unless otherwise stated, the grid resolution used for the reference calculation is $N_\theta=181$ and $N_\phi=360$.

\begin{figure}[t]
    \centering
    \includegraphics[width=0.72\linewidth]{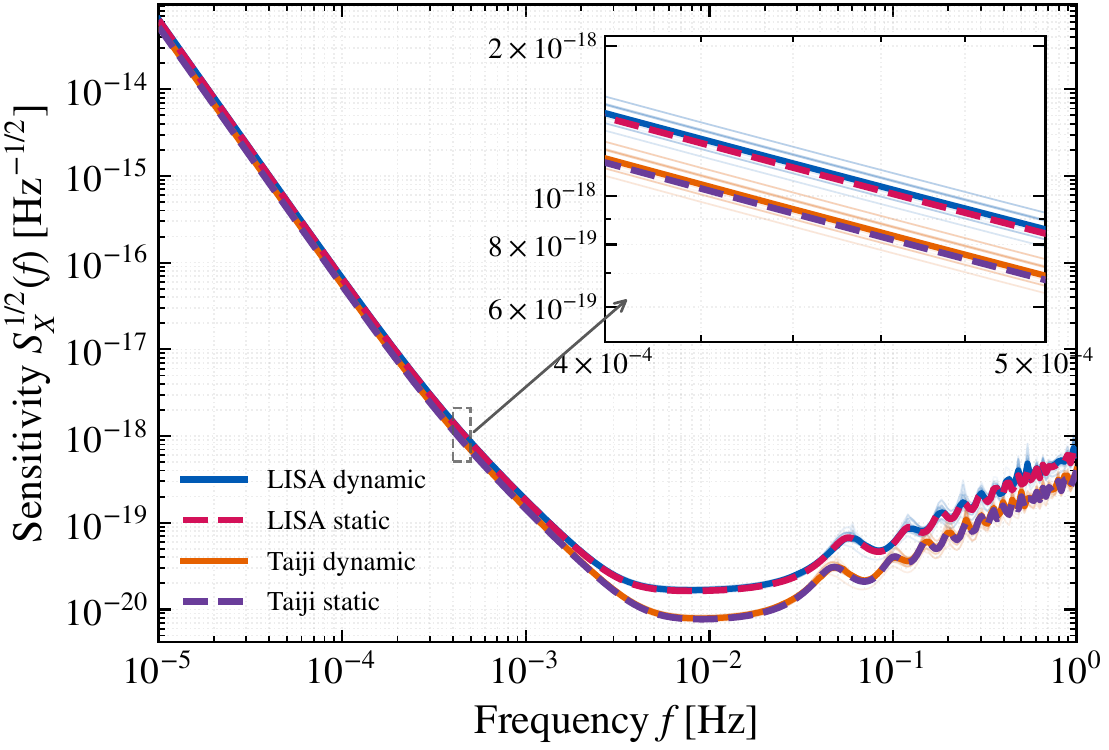}
    \caption{Direction-dependent sensitivity curves in the $X$ channel for dynamic LISA and Taiji for $42$ directions, comparing the dynamic case (solid lines) with the corresponding static sky-averaged references (dashed lines). The inset enlarges the low-frequency region.
    }
    \label{fig:valuable42_sensitivity}
\end{figure}

%\label{subsec:x_dynamic_sensitivities}
\begin{comment}
\begin{figure}[!htbp]
    \centering
    %\includegraphics[width=0.78\linewidth]{fig_x_taiji_lisa_sensitivity.pdf}
    \caption{Sensitivity curves of dynamic constellations in the \(X\) channel for Taiji and LISA, compared with the corresponding static sky-averaged references.}
  \label{fig:x_taiji_lisa_sensitivity}
\end{figure}
\end{comment}
Figure~\ref{fig:valuable42_sensitivity} compares the static reference
with the dynamic sensitivities of the 42 directions for one-year observation time, with
    $\theta\in \left\{ \frac{\pi}{12},\frac{\pi}{6},\frac{\pi}{4}, \frac{\pi}{3},\frac{\pi}{2},\frac{2\pi}{3}, \frac{5\pi}{6} \right\},  
\phi\in \left\{ \frac{\pi}{12},\frac{\pi}{3},\frac{2\pi}{3}, \pi,\frac{4\pi}{3},\frac{5\pi}{3}\right\}$. In both cases,  LISA and Taiji, we notice that the dynamic sensitivity curves (solid lines) scatter around the static one (dashed lines) across the principal millihertz band. 
%Annual orbital averaging narrows the directional spreadaround the static all-sky curve but does not eliminate it. 
And the variation goes to a constant value $\sim 20\%$ at low frequencies but is sizable in the high-frequency regime. Given their similarity, we will concentrate on the Taiji configuration in the remaining discussions.

\subsection{Direction Dependence of the Sensitivity }
\label{subsec:x_low_frequency_plateaus}
\begin{figure}[t]
    \centering
    \includegraphics[width=0.72\linewidth]{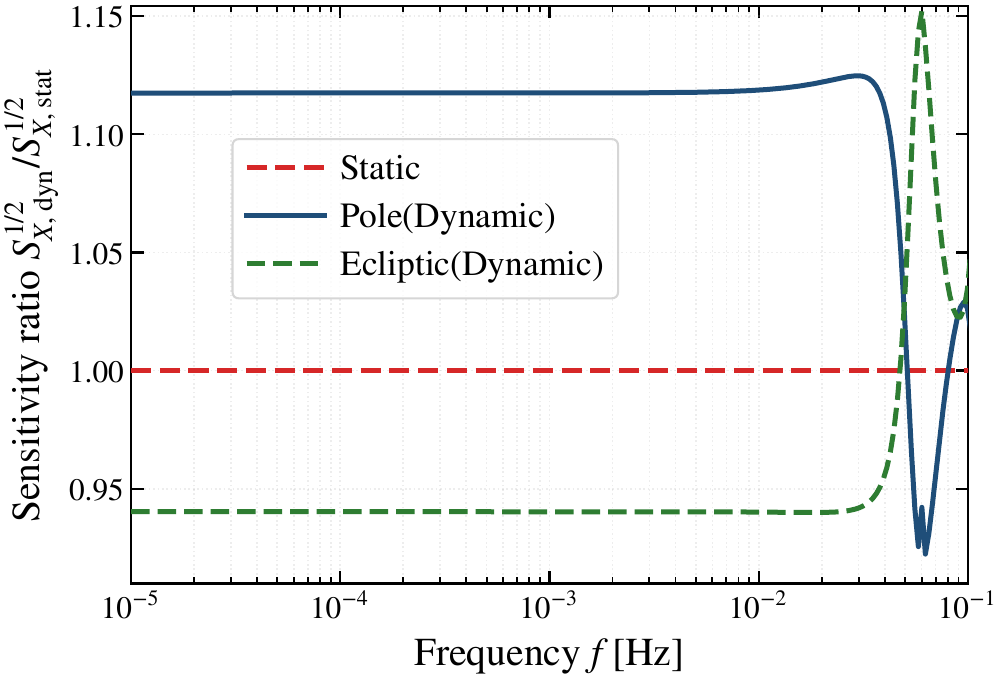}
    \caption{ Ratios relative to the static sky-averaged sensitivity curve. The red line denotes the normalized static reference, the blue curve a pole direction, and the green dashed curve the ecliptic-plane direction \((\theta,\phi)=(\pi/2,\pi/3)\). Smaller values mean better sensitivities. }
    \label{fig:sensitivity_ratio_allsky_pole_ecliptic}
\end{figure}

To quantify the directional difference between the dynamic 
sensitivity and the static equal-arm sky-averaged reference, we
define the ratio
\begin{equation}
\mathcal{Q}_{X}(f,\hat{\Omega})
=
\frac{
S_{X,\rm dyn}^{1/2}(f,\hat{\Omega})
}{
S_{X,\rm stat}^{1/2}(f)
}\simeq \left[
\frac{
\overline{\mathcal{R}}^{\rm stat}_{X}(f)
}{
\overline{\mathcal{R}}^{\rm dyn}_X(f,\hat{\Omega})
}
\right]^{1/2}
.
\label{eq:sensitivity_ratio_allsky}
\end{equation}
Figure~\ref{fig:sensitivity_ratio_allsky_pole_ecliptic} shows
\(\mathcal{Q}_{X}\) for a pole direction and for the
ecliptic-plane direction
\((\theta,\phi)=(\pi/2,\pi/3)\). Both dynamic curves approach nearly
frequency-independent plateaus in the long-wavelength/low-frequency regime. The nearly constant plateaus follow from the long-wavelength
factorization of the annually averaged dynamic response function,
\begin{equation}
\overline{\mathcal{R}}^{\rm dyn}_X(f,\hat{\Omega})
\simeq
16
\left(
\frac{2\pi f}{c}
\right)^4
C_X^{(0)}(\hat{\Omega}),
\qquad
fL/c\ll1,
\label{eq:result_lf_factorization}
\end{equation}
We calculate the leading angular dependence of the response
\begin{equation}
C_X^{(0)}(\theta,\phi)
=
\frac{3L^4}{16384}
\left[
656
+304\sin^2\theta
-74\sin^4\theta
+81\sin^4\theta
\cos\left(4\phi+\frac{\pi}{3}\right)
\right],
\label{eq:fullsky_response_angular_factor}
\end{equation}
and the corresponding static sky-averaged response,
\begin{equation}
\overline{\mathcal{R}}^{\rm stat}_{X}(f)
\simeq
16
\left(
\frac{2\pi f}{c}
\right)^4
C_{X,\rm stat}^{(0)},\; C_{X,\rm stat}^{(0)} = \frac{3}{20}L^4.
\label{eq:static_lf_factorization}
\end{equation}
The details of the calculations are described in the Appendix~\ref{app:low_frequency_factorization}. Note that the common factor \(f^4\)  would cancel in the ratio and in the long-wavelength regime we have a frequency-independent
\begin{equation}
\mathcal{Q}_{X}(f,\hat{\Omega})
\simeq
\left[
\frac{
C_{X,\rm stat}^{(0)}
}{
C_X^{(0)}(\hat{\Omega})
}
\right]^{1/2}.
\label{eq:low_frequency_sensitivity_ratio}
\end{equation}
For example, two directions, pole and $(\pi/2,\pi/3)$,  give
\begin{equation}
\left[
\frac{
C_{X,\rm stat}^{(0)}
}{
C_X^{(0)}(\mathrm{pole})
}
\right]^{1/2}
=
\frac{1}{\sqrt{0.8008}}
\simeq
1.1175,\; 
\left[
\frac{
C_{X,\rm stat}^{(0)}
}{
C_X^{(0)}(\pi/2,\pi/3)
}
\right]^{1/2}
=
\frac{1}{\sqrt{1.131}}
\simeq
0.9403,
\label{eq:ecliptic_geometrical_sensitivity_factor}
\end{equation}
which agree with the numerical evaluation show in Fig.~\ref{fig:sensitivity_ratio_allsky_pole_ecliptic} at low frequency $f\lesssim 0.01$~Hz.

\begin{comment}
The pole direction numerical curve is expected to lie close to, but not
exactly at, the analytic polar value because the plotted direction
has a small nonzero colatitude. 
The polar direction has a weaker response and therefore a larger strain-noise amplitude, whereas the
selected ecliptic direction has a stronger response and therefore a
smaller strain-noise amplitude. 
\end{comment}

%\subsection{Sky-Map of the Sensitivity at $3$~mHz}
%\label{subsec:x_full_sky_distribution}

\begin{figure}[t]
    \centering
    \includegraphics[width=0.8\linewidth]
    {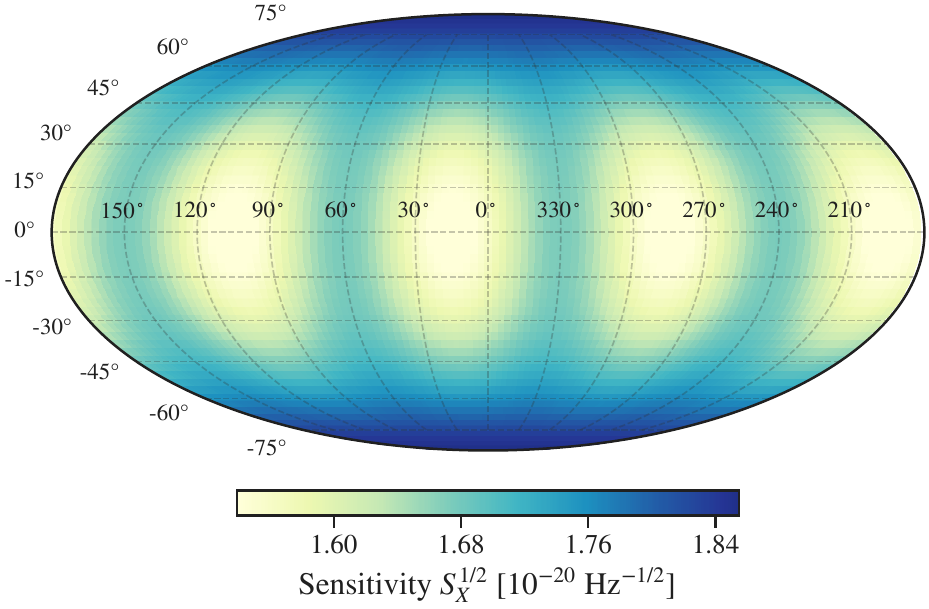}
    \caption{Directional dependence of sensitivity for dynamic Taiji at \(f=3\,\mathrm{mHz}\), shown in ecliptic coordinates. Smaller values correspond to better strain sensitivity. }
    \label{fig:sky_sensitivity_map}
\end{figure}

In Fig.~\ref{fig:sky_sensitivity_map} we plot the dynamic sensitivity over the full sky at \(f=3\,\mathrm{mHz}\), $S_{X,\rm dyn}^{1/2}(f=3\,\mathrm{mHz},\hat{\Omega})$. Note that lower strain values indicate better sensitivity. We therefore conclude that the triangular constellation is more sensitive to sources located in the directions near the ecliptic plane, while its sensitivity in the polar direction is reduced by about $20\%$. We also observe the quadrant pattern in the directional dependence, which arises from the last term $\cos\left({4\phi+\pi/3}\right)$ in Eq.~\ref{eq:fullsky_response_angular_factor}, partially reveals the nature of GWs from quadrupole emission.
%Annual averaging smooths the instantaneous antenna pattern but leaves a coherent large-scale angular structure. The dominant variation is with ecliptic latitude, while the longitude dependence is weaker and is most pronounced near the ecliptic plane.

The above results suggest that the sensitivity would vary about $20\%$ in the low-frequency regime and even more at high frequencies. Because the amplitudes of the GWs are inversely proportional to the distance of the sources $r$ and the volume goes as $r^3$ in low-redshift universe, the number of detectable GW sources would have a strong directional dependence, up to $70\%$ in the low-frequency regime and much larger at high frequencies. Therefore, for a precise estimation of the number of sources in astronomy and parameter inference, one shall use the dynamic and direction-dependent sensitivity curves. The framework for dynamic sensitivity can also be extended to new physics searches~\cite{Pierce:2018xmy, Morisaki:2018htj, Grote:2019uvn, Miller:2023kkd, Yu:2023iog,Yao:2024fie, Yao:2024hap, Xu:2025rfv, Gue:2024txz, Yao:2025vgy, Yao:2025wfd, Liu:2025hwn, Zhang:2025fck, Chen:2026mef}.

\begin{comment}
Because the plotted quantity is the amplitude sensitivity rather than
the response, its leading angular dependence is
\begin{equation}
S_{X,\rm dyn}^{1/2}
\propto
\left[
C_X^{(0)}(\theta,\phi)
\right]^{-1/2}.
\label{eq:fullsky_sensitivity_angular_factor}
\end{equation}
The first three terms in
Eq.~\eqref{eq:fullsky_response_angular_factor} determine the dominant
latitude dependence, while the final term produces a weaker fourfold
longitude modulation. Directions with a larger response coefficient
therefore appear as regions of smaller, and hence better, amplitude
sensitivity in Fig.~\ref{fig:sky_sensitivity_map}. Conversely,
directions with a smaller response coefficient have a larger
strain-noise amplitude.
\end{comment}

\subsection{Sensitivity Curves for Other Interferometric Channels}
\label{subsec:channel_definitions}

\begin{figure}[t] 
    \centering 
    \includegraphics[ width=0.8\linewidth ]{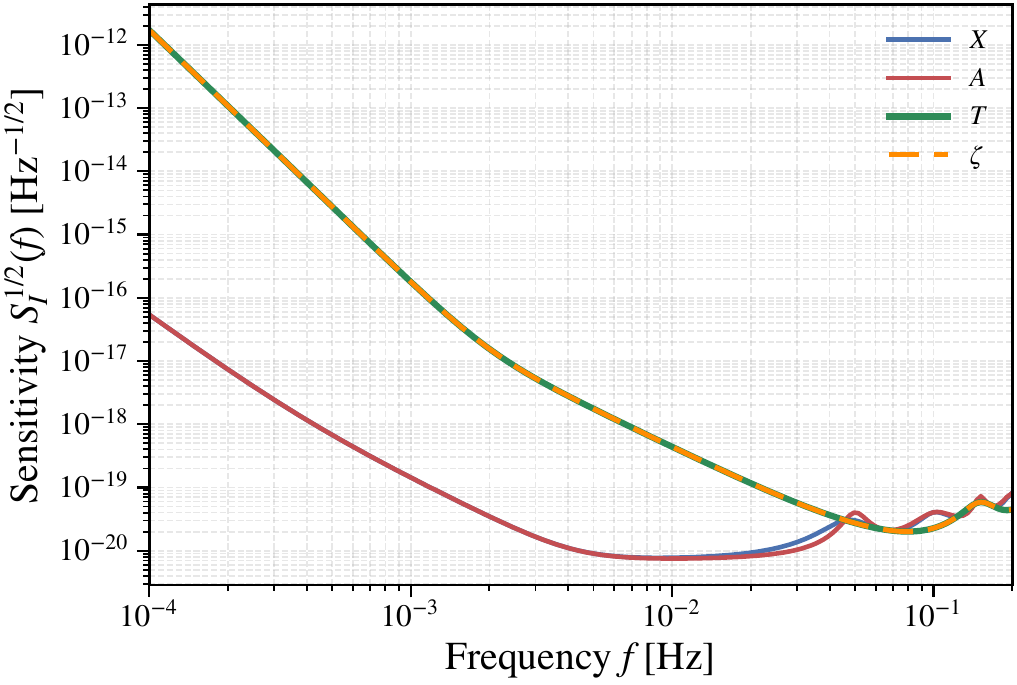}
    \caption{ Static equal-arm sky-averaged amplitude sensitivities of the \(X,A,T,\zeta\) channels, constructed according to Eq.~\eqref{eq:all_sky_sensitivity}. In the symmetric equal-arm limit, the \(A\) and \(E\)-channel sensitivities coincide, so only \(A\) is shown. }\label{fig:static_allsky_channel_sensitivity}
\end{figure}

In this section, we extend the dynamic sensitivity curves for $X$ to other typical interferometric channels. The three Michelson channels $X,Y,Z$ are commonly recombined into three optimal channels $A,E,T$~\cite{Prince:2002hp},
\begin{comment}
This transformation diagonalizes the noise covariance in the symmetric equal-noise limit and separates the Michelson-like science channels from the symmetrized Sagnac-like channel.  The $A$ and $E$ channels carry the main gravitational-wave response at low frequencies, whereas the $T$ channel has a suppressed response in the long-wavelength regime and behaves approximately as a null channel.  In the time-dependent unequal-arm calculation below, the exact noise orthogonality is not assumed; instead, we use the same linear transformation as a conventional channel basis for comparing the dynamic responses.    
\end{comment}
\begin{align}
A = \frac{1}{\sqrt{2}}(Z-X), \;
E = \frac{1}{\sqrt{6}}(X-2Y+Z), \;
T = \frac{1}{\sqrt{3}}(X+Y+Z). \label{eq:T_channel}
\end{align}
In the static and equal-arm configuration, the above three channels are orthogonal. $A$ and $E$ have the same sensitivity, and $T$ acts as a nearly null channel.
We also consider the first-generation symmetrized-Sagnac variable $\zeta$,
%With the link convention that $y_{ij}$ is received at spacecraft $i$ from spacecraft $j$, it is defined in the frozen-arm approximation as
\begin{equation}
\zeta
=
\mathcal{D}_{23}\left(y_{12}-y_{13}\right)
+
\mathcal{D}_{31}\left(y_{23}-y_{21}\right)
+
\mathcal{D}_{12}\left(y_{31}-y_{32}\right).
\label{eq:zeta_channel}
\end{equation}

Fig.~\ref{fig:static_allsky_channel_sensitivity} shows the static equal-arm sensitivities of $X$, $A$, $T$, and $\zeta$ channels, averaging various directions. As shown $T$ and $\zeta$ have much worse sensitivity than other two. However, as we shall show in the dynamic case $T$ has the same sensitivity as $A$ and $X$ in the low-frequency regime. 

\begin{comment}
\rem{,} The static curves separate into two qualitatively different groups. The \(X\) and \(A\) variables retain their leading Michelson-like gravitational-wave response in the long-wavelength regime, whereas the \(T\) and \(\zeta\) responses are more strongly suppressed by cancellations within their respective TDI combinations. The ordering of the sensitivity curves is not determined by the responses alone, because each channel must be combined with its own instrumental-noise transfer function. Nevertheless, the relatively poor low-frequency sensitivities of \(T\) and \(\zeta\) reflect their strongly suppressed signal responses in this regime.
\end{comment}

\begin{figure}[t]
    \centering   
    \includegraphics[width=0.48\linewidth, height=0.38\linewidth]{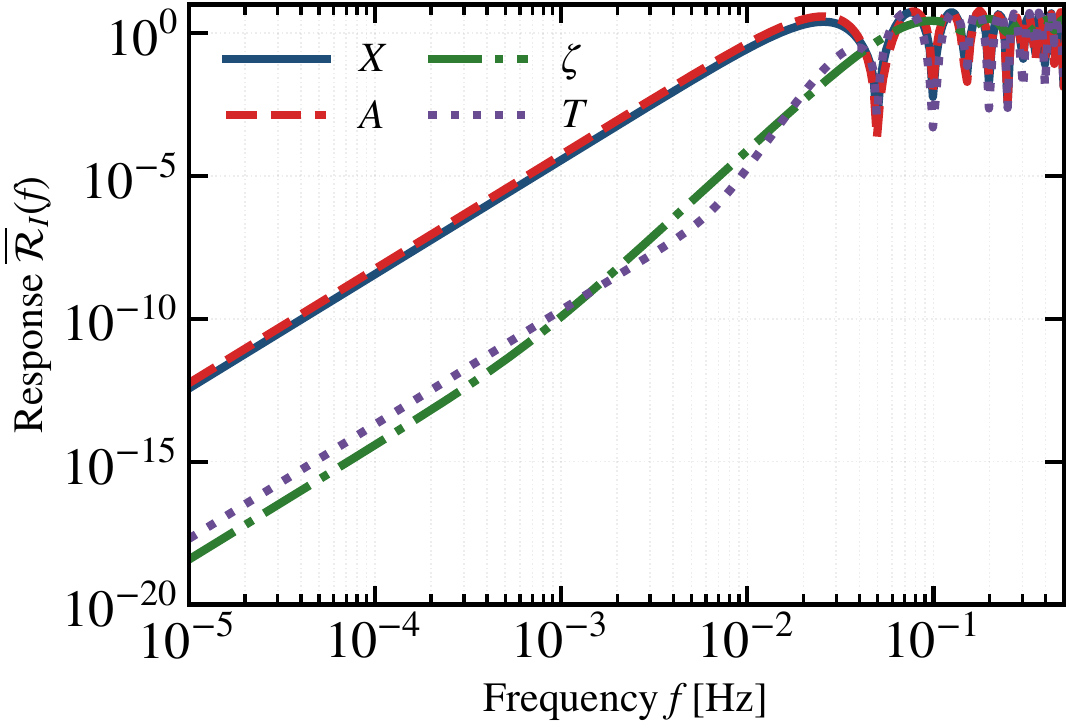}
    \hspace{0.3cm}
    \includegraphics[width=0.48\linewidth, height=0.38\linewidth]{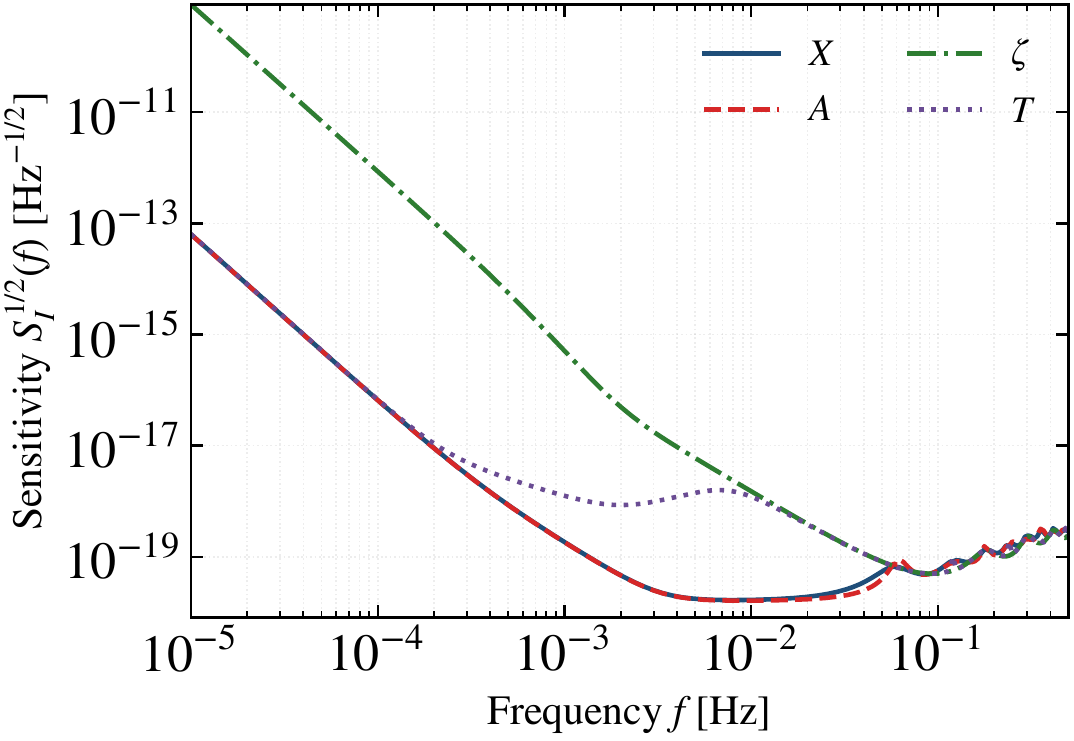}
    \caption{Dynamic responses and amplitude sensitivities of the \(X,A,\zeta,T\) channels for \((\theta,\phi)=(\pi/4,\pi/3)\). The left panel shows the response function \({\mathcal{R}}_I(f)\), and the right panel shows the corresponding sensitivity curves \(S_I^{1/2}(f)\).}
    \label{fig:dyn_channels}
\end{figure}

\begin{figure}[t]
    \centering

    \includegraphics[
        height=0.22\textheight,
        keepaspectratio
    ]{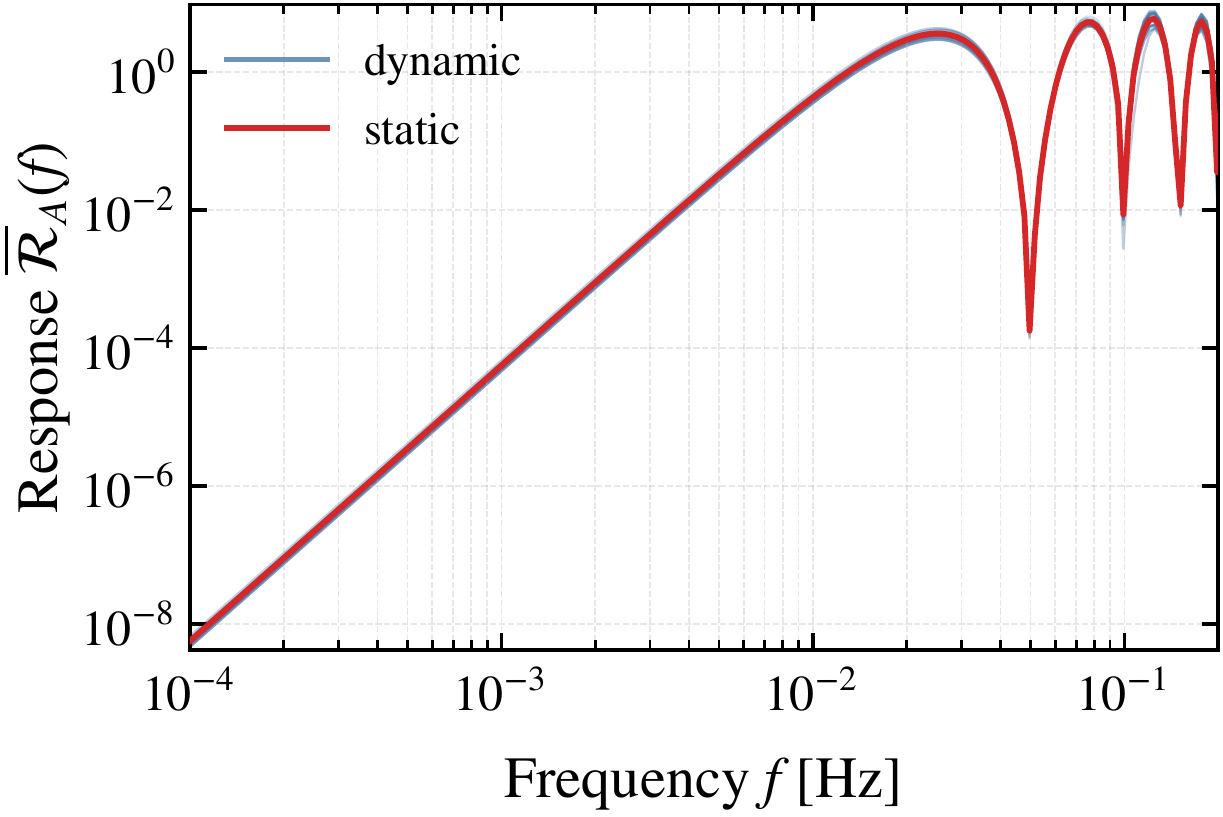}
    \includegraphics[
        height=0.22\textheight,
        keepaspectratio
    ]{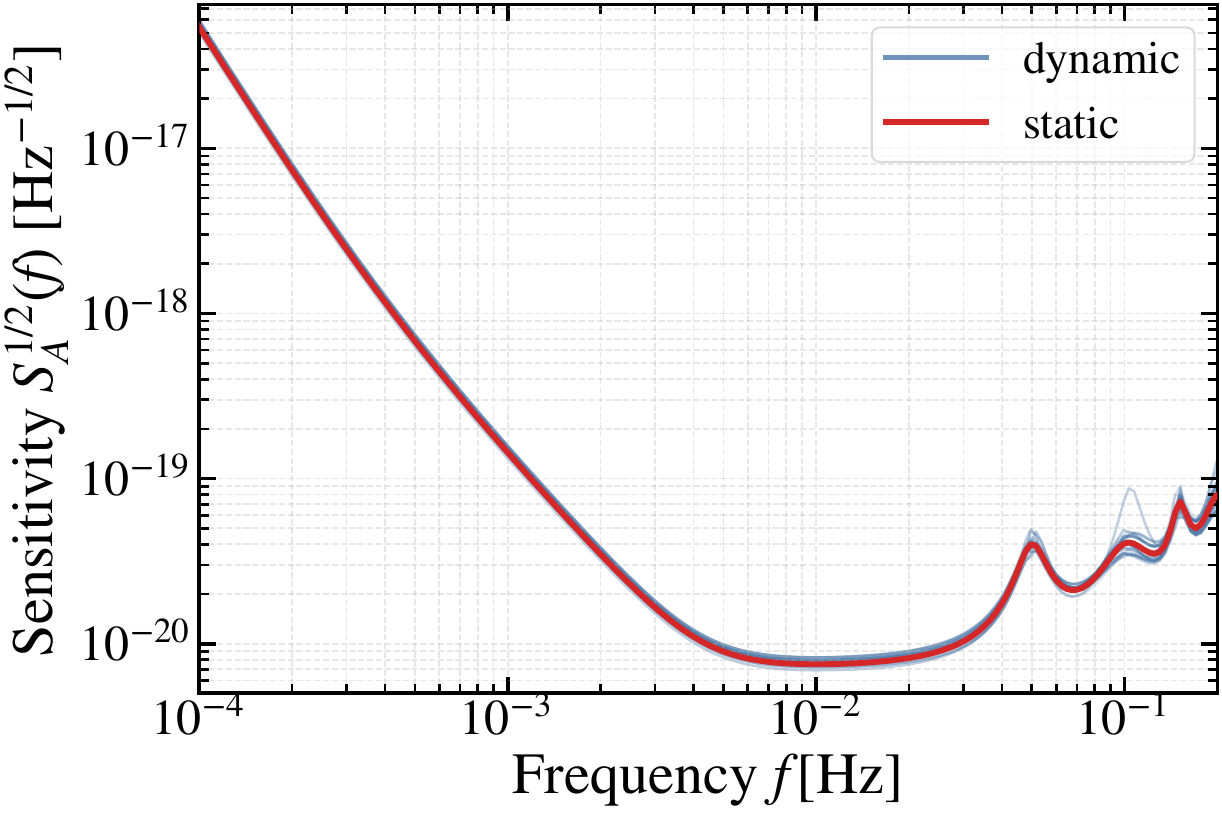}

    \includegraphics[
        height=0.22\textheight,
        keepaspectratio
    ]{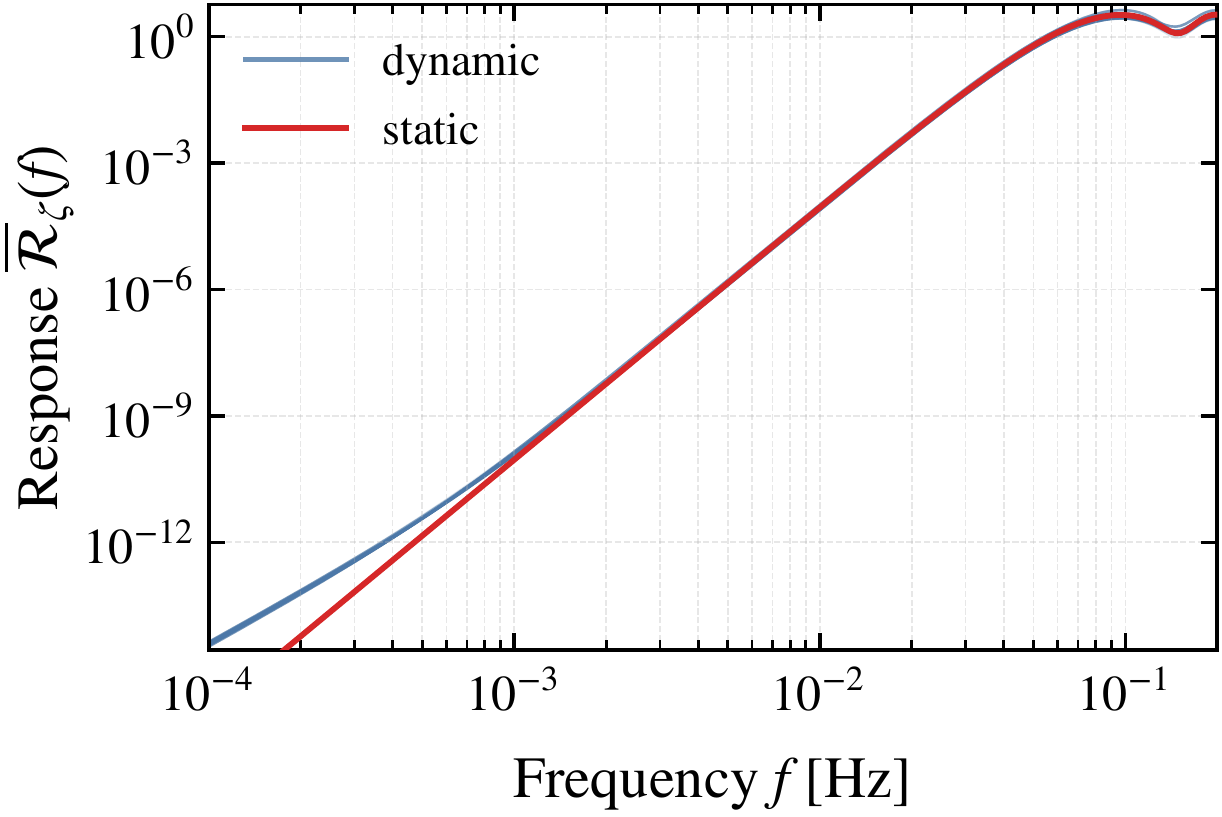}
    \includegraphics[
         height=0.22\textheight,
        keepaspectratio
    ]{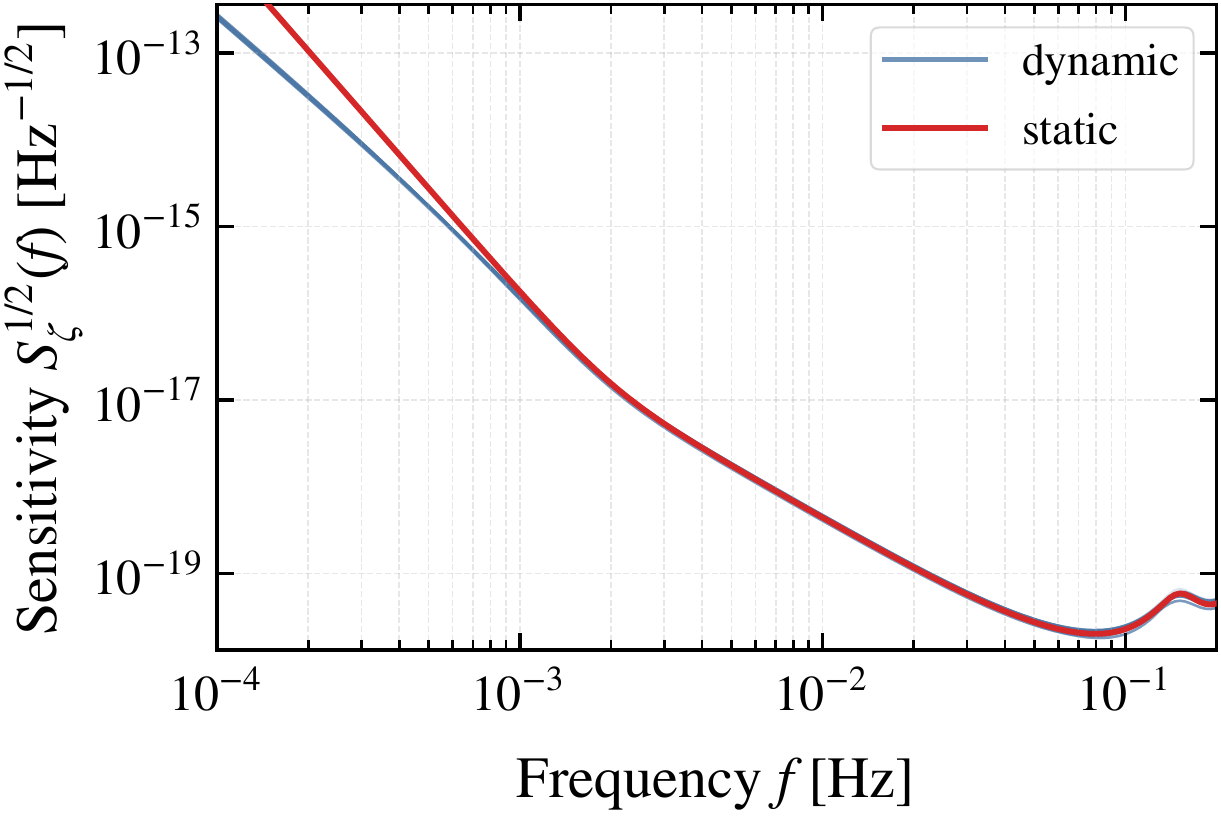} 

    \includegraphics[
        height=0.22\textheight,
        keepaspectratio
    ]{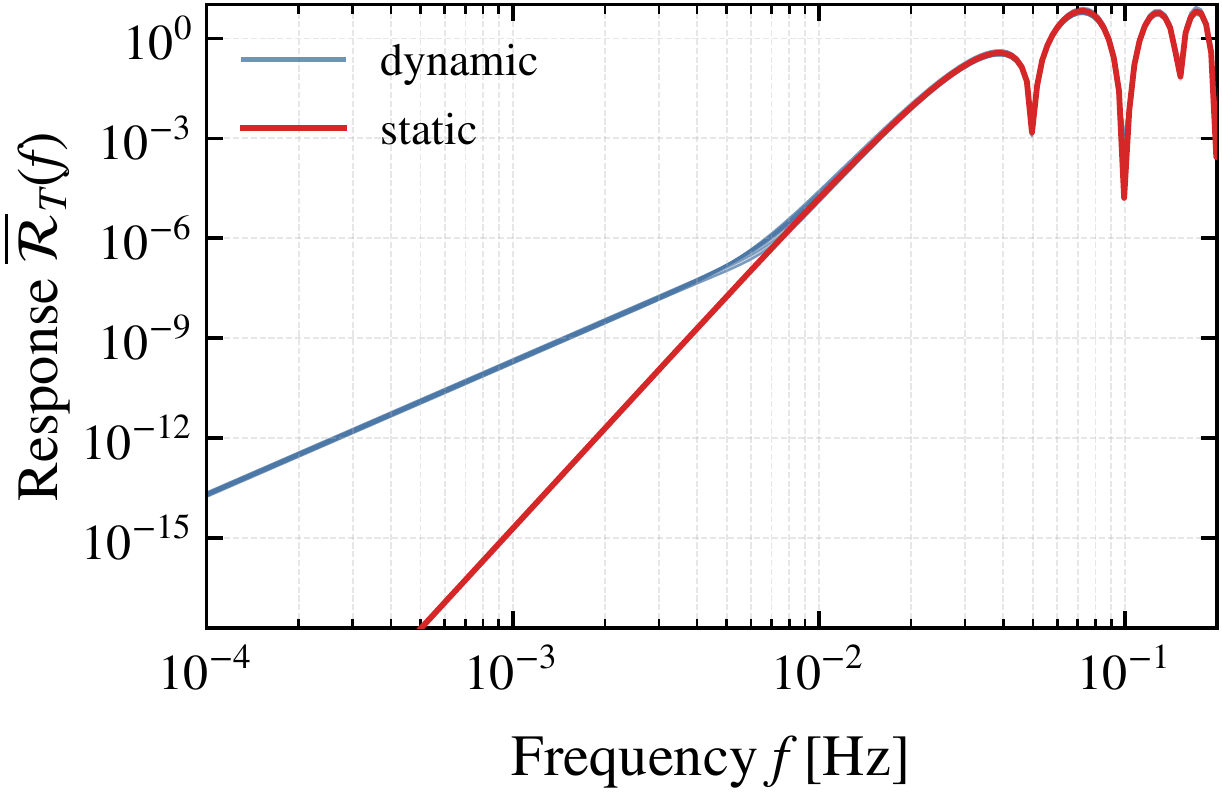}
    \includegraphics[
        height=0.22\textheight,
        keepaspectratio
    ]{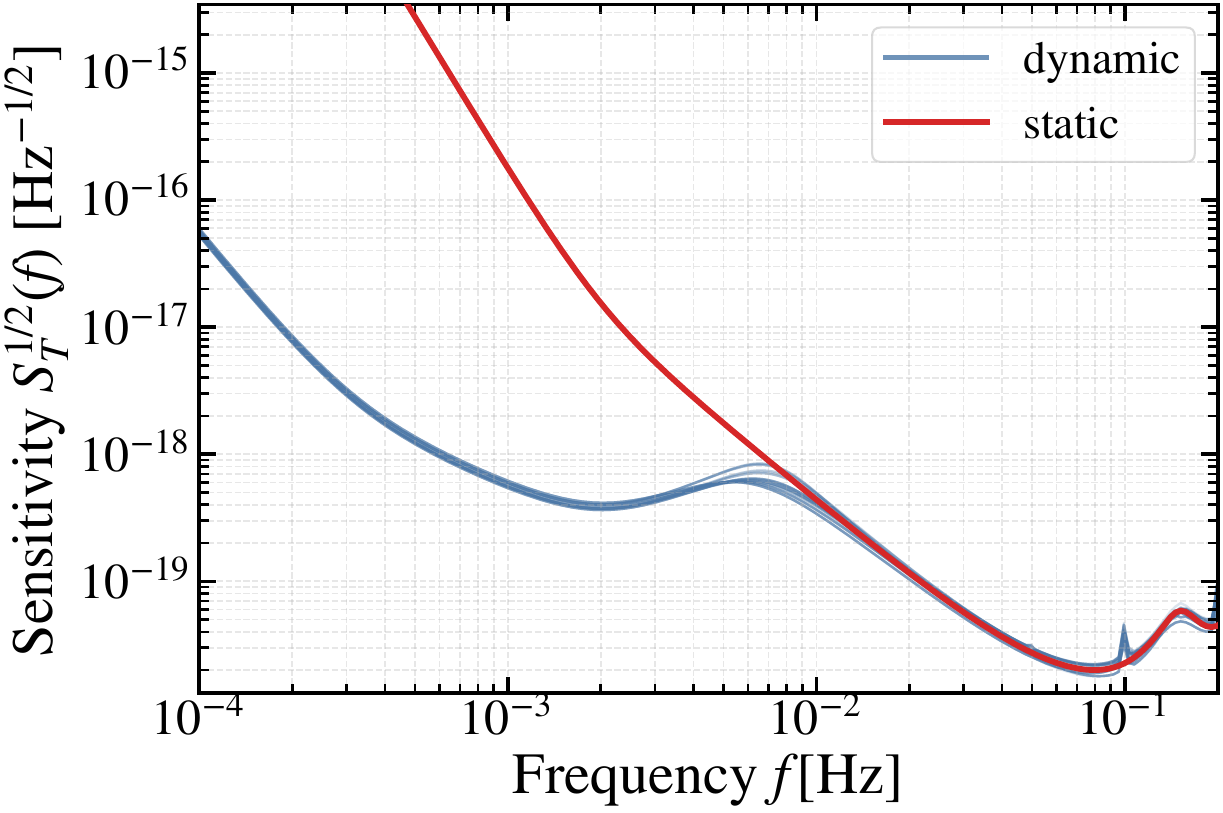}

    \caption{
    Dynamic responses and sensitivities for 42 sky directions,
    compared with the corresponding static equal-arm sky-averaged
    responses for the \(A,\zeta,T\) channels, from top to bottom.
    }
    \label{fig:channel_response_compare}
\end{figure}

In Fig.~\ref{fig:dyn_channels} we show the dynamic response and sensitivity curves of the \(X,A,\zeta,T\) channels for the representative direction \((\theta,\phi)=(\pi/4,\pi/3)\). The responses functions of \(X\) and \(A\)  exhibit almost the same low-frequency scaling and their sensitivity curves only differ from their static ones within $20\%$. However, \(T\) and \(\zeta\) change dramatically in low-frequency regime, see Fig.~\ref{fig:channel_response_compare} of comparison to the static case. For the channel $T$, the reason is that in the dynamic case of the unequal-arms, $A, E$ and $T$ are not orthogonal even more~\cite{Adams:2010vc, Wang:2020fwa}. $T$ gets a modification in the response function and eventually shares the same sensitivity as $A$ and $E$. For the $\zeta$ channel, the modification in the response function is only slightly affected, therefore, its sensitivity is still much worse than the other three, allowing it still as the noise monitor channel~\cite{Muratore:2021uqj, Hartwig:2021mzw}.

\begin{comment}
and \(\zeta\) responses are more strongly suppressed because their leading contributions cancel in the equal-arm long-wavelength limit. The ordering of the response curves does not directly determine the ordering of the sensitivity curves. Each response in the upper panel of Fig.~\ref{fig:dyn_channels} is combined with the instrumental-noise PSD of the same TDI observable to construct the corresponding sensitivity in the lower panel. The two annual sensitivity prescriptions introduced above are indistinguishable on the scale of the figure.
\end{comment}

\section{Summary}
\label{sec:summary}
We have investigated the direction-dependent responses and sensitivity curves of dynamic Taiji and LISA, using an adiabatically evolving unequal-arm triangular constellation with a heliocentric orbit, and compared with the corresponding static equal-arm sky-averaged reference. We have analytically derived the leading angular dependence of the sensitivity in the low-frequency/long-wavelength regime ($\lesssim 0.01$~Hz) and found the variation is about $20\%$ in the dynamic case. This would lead to a $70\%$ change in the estimation of the number of GW sources from different directions. At higher frequencies, the variation could be even greater. These results suggest that, for a precise estimation of the number of GW sources and parameter inference of binary systems in astrophysics, one shall use the dynamic and direction-dependent sensitivity curves.

\begin{acknowledgments}
This work is partly supported by the National Key Research and Development Program of China (Grant No.~2021YFC2201901), 
the National Natural Science Foundation (Grant No.12547104), and the Fundamental Research Funds for the Central Universities. 
\end{acknowledgments}

\appendix
\section{Instrumental noises}
\label{app:noise_model}

\subsection{Power Spectral Density (PSD) of noises}

The optical metrology noise and test-mass acceleration noise are
modeled as~\cite{Larson2000Sensitivity,BayleHartwig2023LISAInstrument}
\begin{align}
    S_{\rm oms}^{1/2}(f)
    &=
        15\times10^{-12}\,
        \frac{2\pi f}{c}
        \sqrt{
            1+
            \left(
                \frac{2\times10^{-3}\,\mathrm{Hz}}{f}
            \right)^4
        }\frac{\mathrm{m}}{\sqrt{\mathrm{Hz}}},
    \label{eq:S_oms}
    \\
    S_{\rm acc}^{1/2}(f)
    &=
        \frac{3\times10^{-15}}{2\pi f c}
        \sqrt{
            \left[
                1+
                \left(
                    \frac{0.4\times10^{-3}\,\mathrm{Hz}}{f}
                \right)^2
            \right]
            \left[
                1+
                \left(
                    \frac{f}{8\times10^{-3}\,\mathrm{Hz}}
                \right)^4
            \right]
        }\frac{\mathrm{m/s^2}}{\sqrt{\mathrm{Hz}}}.
    \label{eq:S_acc}
\end{align}
The numerical coefficient \(15\times10^{-12}\)
in Eq.~\eqref{eq:S_oms} corresponds to the LISA optical metrology
noise level. For Taiji, we replace it by
\(8\times10^{-12}\), while keeping the
same functional form of the OMS noise spectrum.

For the fixed equal-arm compact $X$ variable, the corresponding static
noise PSD is
\begin{equation}
    P_X^{\rm stat}(f)
    =
    16
    \sin^2\left(
        \frac{2\pi fL}{c}
    \right)
    \left[
        S_{\rm oms}(f)
        +
        \left(
            3+
            \cos\frac{4\pi fL}{c}
        \right)
        S_{\rm acc}(f)
    \right].
    \label{eq:X_noise_psd}
\end{equation}
Here $L$ is the fiducial arm length of the constellation. This expression is combined with the static sky-averaged
response to construct the conventional static equal-arm sensitivity for $X$-channel. The static noise PSDs of the $A$, $E$, and
$T$ channels are obtained from the equal-arm $X,Y,Z$ noise covariance,
while the static $\zeta$ noise PSD is constructed from the same
symmetrized-Sagnac combination used for its response.

\subsection{Unequal-arm noise construction}

In the dynamic case, the noise spectra in each interferometric channel are constructed
directly from the single-link optical metrology and acceleration-noise
PSDs in Eqs.~\eqref{eq:S_oms} and \eqref{eq:S_acc}, using the same
adiabatic unequal-arm TDI combinations as those used for the
gravitational-wave response.

For each orbital day, we compute the daily mean arm lengths
$L_{12}$, $L_{13}$, and $L_{23}$ from the time-dependent spacecraft
positions. In the present adiabatic approximation, the directed
light-travel times are taken to be reciprocal,
\begin{equation}
    L_{21}=L_{12},
    \qquad
    L_{31}=L_{13},
    \qquad
    L_{32}=L_{23}.
    \label{eq:reciprocal_arms_noise}
\end{equation}

The corresponding delay factors are
\begin{equation}
    D_{ij}(f)
    =
    \exp\left(
        \frac{2\pi i fL_{ij}}{c}
    \right).
    \label{eq:noise_delay_factor}
\end{equation}
The positive sign follows from the Fourier convention
$e^{-2\pi i f t}$ used in both the response and noise calculations.

For a one-way Doppler link $ij$, the instrumental-noise contribution
is modeled as
\begin{equation}
    N_{ij}
    =
    n^{\rm oms}_{ij}
    +
    n^{\rm acc}_{ij}
    +
    D_{ij}n^{\rm acc}_{ji},
    \label{eq:link_noise_model}
\end{equation}
where $n^{\rm oms}_{ij}$ and $n^{\rm acc}_{ij}$ denote independent
optical metrology and test-mass acceleration noises associated with
the directed link. The round-trip noise combinations are
\begin{equation}
    M_{ij}
    =
    N_{ij}
    +
    D_{ij}N_{ji}.
    \label{eq:round_trip_noise}
\end{equation}

Explicitly, the noises of the three Michelson variables are
\begin{align}
    X
    &=
    \left(
        1-D_{13}D_{31}
    \right)M_{12}
    +
    \left(
        D_{12}D_{21}-1
    \right)M_{13},
    \label{eq:X_noise_combination}
    \\
    Y
    &=
    \left(
        1-D_{21}D_{12}
    \right)M_{23}
    +
    \left(
        D_{23}D_{32}-1
    \right)M_{21},
    \label{eq:Y_noise_combination}
    \\
    Z
    &=
    \left(
        1-D_{32}D_{23}
    \right)M_{31}
    +
    \left(
        D_{31}D_{13}-1
    \right)M_{32}.
    \label{eq:Z_noise_combination}
\end{align}
This construction retains the frequency-dependent transfer
coefficients of each independent underlying OMS and acceleration-noise
source.

Equivalently, each Michelson channel can be written as a linear
combination of independent noise sources,
\begin{equation}
    I(f)
    =
    \sum_q
    C_I^q(f)n_q(f),
    \qquad
    I\in\{X,Y,Z\},
    \label{eq:channel_noise_linear_combination}
\end{equation}
where $q$ labels the independent OMS and acceleration-noise components,
and $C_I^q(f)$ is the corresponding TDI transfer coefficient.

The noise cross-spectral density matrix is
\begin{equation}
    P_{IJ}(f)
    =
    \sum_q
    C_I^q(f)
    \left[
        C_J^q(f)
    \right]^*
    S_q(f),
    \qquad
    I,J\in\{X,Y,Z\},
    \label{eq:xyz_noise_covariance}
\end{equation}
with $S_q=S_{\rm oms}$ or $S_{\rm acc}$ depending on the type of
noise source.

The $A$, $E$, and $T$ noise covariance is obtained by applying
\begin{equation}
    \begin{pmatrix}
        A\\
        E\\
        T
    \end{pmatrix}
    =
    \begin{pmatrix}
        -1/\sqrt{2} & 0 & 1/\sqrt{2}\\
        1/\sqrt{6} & -2/\sqrt{6} & 1/\sqrt{6}\\
        1/\sqrt{3} & 1/\sqrt{3} & 1/\sqrt{3}
    \end{pmatrix}
    \begin{pmatrix}
        X\\
        Y\\
        Z
    \end{pmatrix}.
    \label{eq:aet_noise_transform}
\end{equation}
The diagonal elements of the transformed covariance matrix give
$P_A(f)$, $P_E(f)$, and $P_T(f)$, while $P_X(f)$ is obtained from
the $XX$ element of Eq.~\eqref{eq:xyz_noise_covariance}.

For the standard symmetrized-Sagnac channel, the instrumental-noise
combination is
\begin{equation}
    N_\zeta
    =
    D_{23}
    \left(
        N_{12}-N_{13}
    \right)
    +
    D_{31}
    \left(
        N_{23}-N_{21}
    \right)
    +
    D_{12}
    \left(
        N_{31}-N_{32}
    \right).
    \label{eq:zeta_noise_combination}
\end{equation}
Its noise PSD is obtained by expanding
Eq.~\eqref{eq:zeta_noise_combination} in the independent OMS and
acceleration-noise sources and summing their squared transfer
coefficients, in the same manner as
Eq.~\eqref{eq:xyz_noise_covariance}. The response and noise of $\zeta$ 
are therefore evaluated with identical link, delay-phase, and
arm-length conventions.

\section{Response function at low frequencies}
\label{app:low_frequency_factorization}

This section presents the derivation of the nearly constant low-frequency response-ratio plateaus found in Sec.~\ref{sec:results}.  
The derivation follows the same adiabatic, reciprocal-arm, commuting-delay approximation used in the numerical implementation.  
It is therefore a low-frequency derivation for the compact TDI model used here, not a second-generation TDI treatment with noncommuting delays~\cite{TintoDhurandhar2005TDI}.

We start from the Michelson channel $X$,
\begin{equation}
X=
\left(1-\mathcal{D}_{13}\mathcal{D}_{31}\right)
\eta_{12}
-
\left(1-\mathcal{D}_{12}\mathcal{D}_{21}\right)\eta_{13},
\label{eq:app_X_compact}
\end{equation}
where
\begin{equation}
\eta_{12}=y_{12}+\mathcal{D}_{12}y_{21},
\qquad
\eta_{13}=y_{13}+\mathcal{D}_{13}y_{31}.
\end{equation}
Under the reciprocal-arm approximation,
\begin{equation}
\mathcal{D}_{ij}\mathcal{D}_{ji}\simeq D_{ij}^2,
\qquad
D_{ij}=\exp\left(+2\pi i f L_{ij}/c\right).
\end{equation}

Define
\begin{equation}
x_{ij}(t)=\frac{2\pi f L_{ij}(t)}{c}.
\end{equation}
In the long-wavelength limit \(x_{ij}\ll 1\), the one-link gravitational-wave response can be expanded as
\begin{equation}
y_{ij}^{A}(f,t,\hat{\Omega})
\simeq
 i x_{ij}(t)\,
d_{ij}^{A}(t,\hat{\Omega})\,h_A(f)
+
\mathcal{O}(x_{ij}^2),
\label{eq:app_onelink_lf}
\end{equation}
where
\begin{equation}
d_{ij}^{A}(t,\hat{\Omega})
=
\frac{1}{2}
\hat n_{ij}^{a}(t)\hat n_{ij}^{b}(t)
e^{A}_{ab}(\hat{\Omega})
\label{eq:app_dij_def}
\end{equation}
is the geometrical antenna projection in the low-frequency limit.  
The apparent denominator in the exact one-link response cancels against the first-order expansion of the propagation phase, leaving the usual quadrupolar projection factor.

Because \(D_{ij}=1+\mathcal{O}(x_{ij})\) and \(d_{ji}^{A}=d_{ij}^{A}\), the round-trip combinations become
\begin{equation}
\eta_{12}^{A}
\simeq
2i x_{12}d_{12}^{A}h_A
+
\mathcal{O}(x^2),
\qquad
\eta_{13}^{A}
\simeq
2i x_{13}d_{13}^{A}h_A
+
\mathcal{O}(x^2).
\label{eq:app_eta_lf}
\end{equation}
The delay prefactor similarly gives
\begin{equation}
1-D_{ij}^2
\simeq
-2i x_{ij}
+
\mathcal{O}(x_{ij}^2).
\label{eq:app_delay_lf}
\end{equation}
Substituting Eqs.~\eqref{eq:app_eta_lf} and \eqref{eq:app_delay_lf} into Eq.~\eqref{eq:app_X_compact}, one obtains the leading-order TDI response
\begin{equation}
X^A(f,t,\hat{\Omega})
\simeq
4
\left(\frac{2\pi f}{c}\right)^2
L_{12}(t)L_{13}(t)
\left[
d_{12}^{A}(t,\hat{\Omega})
-
d_{13}^{A}(t,\hat{\Omega})
\right]h_A(f).
\label{eq:app_X_lf}
\end{equation}
Thus the amplitude of $X$ in frequency domain scales as \(f^2\) in the low-frequency limit, and the response function scales as \(f^4\).

The polarization-averaged response is
\begin{equation}
\mathcal{R}_X(f,t,\hat{\Omega})
=
\frac{1}{2}
\sum_{A=+,\times}
\left|
\frac{X^A(f,t,\hat{\Omega})}{h_A(f)}
\right|^2 .
\end{equation}
Using Eq.~\eqref{eq:app_X_lf}, we find
\begin{equation}
\mathcal{R}_X(f,t,\hat{\Omega})
\simeq
16
\left(\frac{2\pi f}{c}\right)^4
L_{12}^2(t)L_{13}^2(t)
\frac{1}{2}
\sum_A
\left|
d_{12}^{A}(t,\hat{\Omega})
-
d_{13}^{A}(t,\hat{\Omega})
\right|^2 .
\label{eq:app_RX_lf_time}
\end{equation}

Let \(\alpha=2\pi t/T_{\rm yr}\) be the orbital phase over one year. 
Then
\begin{equation}
\frac{1}{T_{\rm yr}}\int_0^{T_{\rm yr}} dt
=
\frac{1}{2\pi}\int_0^{2\pi}d\alpha .
\end{equation}
The annually averaged response becomes
\begin{equation}
\overline{\mathcal{R}}_X(f,\hat{\Omega})
\simeq
16
\left(\frac{2\pi f}{c}\right)^4
C_X(\hat{\Omega}),
\label{eq:app_RX_factorized}
\end{equation}
where the direction-dependent geometrical coefficient is
\begin{equation}
C_X(\hat{\Omega})
=
\frac{1}{2\pi}
\int_0^{2\pi}
d\alpha\,
L_{12}^2(\alpha)L_{13}^2(\alpha)
\frac{1}{2}
\sum_A
\left|
d_{12}^{A}(\alpha,\hat{\Omega})
-
d_{13}^{A}(\alpha,\hat{\Omega})
\right|^2 .
\label{eq:app_CX_def}
\end{equation}
Eq.~\eqref{eq:app_RX_factorized} shows that the low-frequency response factorizes into a universal \(f^4\) dependence and a direction-dependent coefficient.

For two directions \(\hat{\Omega}_1\) and \(\hat{\Omega}_2\), the low-frequency response ratio is therefore
\begin{equation}
\frac{
\overline{\mathcal{R}}_X(f,\hat{\Omega}_1)
}{
\overline{\mathcal{R}}_X(f,\hat{\Omega}_2)
}
\simeq
\frac{
C_X(\hat{\Omega}_1)
}{
C_X(\hat{\Omega}_2)
},
\qquad
fL/c\ll1.
\label{eq:app_direction_ratio}
\end{equation}
The same argument applies when one of the two quantities is the sky-averaged response. 
The all-sky average is
\begin{equation}
\overline{\mathcal{R}}_{X}^{\rm stat}(f)
=
\frac{1}{4\pi}
\int d\hat{\Omega}\,
\overline{\mathcal{R}}_X(f,\hat{\Omega}).
\end{equation}
Using Eq.~\eqref{eq:app_RX_factorized}, we obtain
\begin{equation}
\overline{\mathcal{R}}_{X}^{\rm stat}(f)
\simeq
16
\left(\frac{2\pi f}{c}\right)^4
C_{X,\rm stat},
\end{equation}
where
\begin{equation}
C_{X,\rm stat}
=
\frac{1}{4\pi}
\int d\hat{\Omega}\,
C_X(\hat{\Omega}).
\end{equation}
Therefore,
\begin{equation}
\frac{
\overline{\mathcal{R}}_X(f,\hat{\Omega})
}{
\overline{\mathcal{R}}_{X}^{\rm stat}(f)
}
\simeq
\frac{
C_X(\hat{\Omega})
}{
C_{X,\rm stat}
},
\qquad
fL/c\ll1.
\label{eq:app_allsky_ratio}
\end{equation}

For two directions, the same annual-averaged dynamic noise PSD appears in both sensitivities.  Their low-frequency ratio is therefore
\begin{equation}
\frac{
S_{X,\rm dyn}^{1/2}(f,\hat{\Omega}_1)
}{
S_{X,\rm dyn}^{1/2}(f,\hat{\Omega}_2)
}
\simeq
\sqrt{
\frac{
C_X(\hat{\Omega}_2)
}{
C_X(\hat{\Omega}_1)
}
}.
\label{eq:app_sens_direction_ratio}
\end{equation}

For the comparison between a dynamic directional sensitivity and the conventional static equal-arm sky-averged sensitivity, the noise PSDs could be different for other channels $I$. The corresponding low-frequency relation is
\begin{equation}
\frac{
S_{I,\rm dyn}^{1/2}(f,\hat{\Omega})
}{
S_{I,\rm stat}^{1/2}(f)
}
\simeq
\left[
\frac{
\overline{P}^{\rm dyn}_I(f)
}{
P_I^{\rm stat}(f)
}
\right]^{1/2}
\sqrt{
\frac{
C_{I,\mathrm{sky}}
}{
C_I(\hat{\Omega})
}
}.
\label{eq:app_sens_allsky_ratio}
\end{equation}
The sensitivity ratio therefore contains both the inverse-square-root response factor and a direction-independent noise-transfer factor.

%\subsection{Explicit angular factor in the rigid-cartwheel limit}
%\label{app:explicit_angular_factor}

In the leading equal-arm rigid-cartwheel limit, the general
low-frequency coefficient \(C_X(\hat{\Omega})\) defined in
Eq.~\eqref{eq:app_CX_def} reduces to the zeroth-order coefficient
\(C_X^{(0)}(\hat{\Omega})\). We now derive a closed-form expression for
\(C_X^{(0)}(\hat{\Omega})\) in this limit.
The full numerical calculation used in the main text retains the \(e^2\) orbital terms and the slow arm-length breathing, whereas the expression derived below keeps only the leading equal-arm cartwheeling geometry.

We first introduce the low-frequency detector tensor for the \(X\) channel,
\begin{equation}
D_X^{ab}(\alpha)
=
\hat n_{12}^{a}(\alpha)\hat n_{12}^{b}(\alpha)
-
\hat n_{13}^{a}(\alpha)\hat n_{13}^{b}(\alpha),
\label{eq:app_detector_tensor_X}
\end{equation}
where \(\alpha\) is the annual orbital phase. 
Then
\begin{equation}
d_{12}^{A}-d_{13}^{A}
=
\frac{1}{2}
D_X^{ab}e^{A}_{ab}(\hat{\Omega}).
\label{eq:app_delta_d_detector_tensor}
\end{equation}
The polarization sum can be written in terms of the transverse projector
\begin{equation}
P_{ab}(\hat{\Omega})=\delta_{ab}-\hat{\Omega}_a\hat{\Omega}_b
\end{equation}
as
\begin{equation}
\sum_{A=+,\times}
e^A_{ab}(\hat{\Omega})e^A_{cd}(\hat{\Omega})
=
P_{ac}P_{bd}
+
P_{ad}P_{bc}
-
P_{ab}P_{cd}.
\label{eq:app_polarization_sum}
\end{equation}
Using Eqs.~\eqref{eq:app_delta_d_detector_tensor} and \eqref{eq:app_polarization_sum}, we get the polarization-averaged geometrical factor
\begin{equation}
\mathcal{G}_X(\alpha,\hat{\Omega})
=
\frac{1}{2}
\sum_A
\left|
d_{12}^{A}(\alpha,\hat{\Omega})
-
d_{13}^{A}(\alpha,\hat{\Omega})
\right|^2
=
\frac{1}{8}
D_X^{ab}(\alpha)
D_X^{cd}(\alpha)
\Lambda_{abcd}(\hat{\Omega}),
\label{eq:app_GX_tensor}
\end{equation}
where
\begin{equation}
\Lambda_{abcd}(\hat{\Omega})
=
P_{ac}P_{bd}
+
P_{ad}P_{bc}
-
P_{ab}P_{cd}.
\label{eq:app_Lambda_def}
\end{equation}

In the leading-order rigid-cartwheel limit, we use $L_{12}=L_{13}=L$ and obtain the angular factor
\begin{equation}
C_X^{(0)}(\hat{\Omega})
=
L^4
\left\langle
\mathcal{G}_X^{(0)}(\alpha,\hat{\Omega})
\right\rangle_{\alpha},
\qquad
\left\langle \cdots \right\rangle_{\alpha}
=
\frac{1}{2\pi}
\int_0^{2\pi}(\cdots)\,d\alpha .
\label{eq:app_CX_rigid_average}
\end{equation}

\begin{comment}
The source-direction and propagation-direction conventions used in
the numerical calculation are
\begin{equation}
    \hat{\mathbf{N}}
    =
    \left(
        \sin\theta\cos\phi,\,
        \sin\theta\sin\phi,\,
        \cos\theta
    \right),
    \qquad
    \hat{\Omega}
    =
    -\hat{\mathbf{N}}.
    \label{eq:app_sky_directions}
\end{equation}
The corresponding sky-fixed transverse basis is
\begin{equation}
    \hat{u}
    =
    \left(
        \cos\theta\cos\phi,\,
        \cos\theta\sin\phi,\,
        -\sin\theta
    \right),
    \qquad
    \hat{v}
    =
    \left(
        -\sin\phi,\,
        \cos\phi,\,
        0
    \right).
    \label{eq:app_uv_sky_convention}
\end{equation}    
\end{comment}

At leading order in the orbital eccentricity, we have the two arm directions
\begin{equation}
\hat n_{12}^{(0)}=
\left(
-\frac14\sin 2\alpha,
1-\frac12\sin^2\alpha,
\frac{\sqrt3}{2}\sin\alpha
\right),
\label{eq:app_n12_leading}
\end{equation}
\begin{equation}
\hat n_{13}^{(0)}=
\left(
\frac14\cos(2\alpha+\pi/6)-\frac{3\sqrt3}{8},
\frac14\sin(2\alpha+\pi/6)+\frac38,
-\frac{\sqrt3}{2}\cos(\alpha+\pi/6)
\right).
\label{eq:app_n13_leading}
\end{equation}

Substituting Eqs.~\eqref{eq:app_n12_leading} and \eqref{eq:app_n13_leading} into Eq.~\eqref{eq:app_GX_tensor} and averaging the resulting trigonometric polynomial over \(\alpha\) removes all nonzero orbital harmonics. 
The remaining closed-form result is
\begin{equation}
C_X^{(0)}(\theta,\phi)
=
\frac{3L^4}{16384}
\left[
656
+304\sin^2\theta
-74\sin^4\theta
+81\sin^4\theta
\cos\left(4\phi+\frac{\pi}{3}\right)
\right].
\label{eq:app_CX_explicit_fullsky}
\end{equation}
 The terms depending only on \(\theta\) determine the dominant latitude dependence, while the longitude dependence appears as a weaker fourfold modulation.

For a polar source, \(\theta=0\), the longitude is irrelevant and
\begin{equation}
C_X^{(0)}(\mathrm{pole})
=
L^4\frac{123}{1024}.
\label{eq:app_CX_pole_explicit}
\end{equation}
For a source on the ecliptic plane, \(\theta=\pi/2\), Eq.~\eqref{eq:app_CX_explicit_fullsky} reduces to
\begin{equation}
C_X^{(0)}\left(\frac{\pi}{2},\phi\right)
=
\frac{3L^4}{16384}
\left[
886
+
81\cos\left(4\phi+\frac{\pi}{3}\right)
\right].
\label{eq:app_CX_ecliptic_explicit}
\end{equation}
The sky average of Eq.~\eqref{eq:app_CX_explicit_fullsky} is
\begin{equation}
C_{X,\rm stat}^{(0)}
=
\frac{1}{4\pi}
\int d\hat{\Omega}\,
C_X^{(0)}(\theta,\phi)
=
\frac{3}{20}L^4,
\label{eq:app_CX_sky_explicit}
\end{equation}
where we used
\begin{equation}
\langle \sin^2\theta\rangle_{\rm sky}=\frac{2}{3},
\qquad
\langle \sin^4\theta\rangle_{\rm sky}=\frac{8}{15},
\qquad
\langle \cos(4\phi+\pi/3)\rangle_{\phi}=0.
\end{equation}
Therefore, the leading-order response plateau of the polar direction relative to the sky-average is
\begin{equation}
\frac{
C_X^{(0)}(\mathrm{pole})
}{
C_{X,\rm stat}^{(0)}
}
=
\frac{123/1024}{3/20}
\simeq
0.8008.
\label{eq:app_pole_allsky_ratio_explicit}
\end{equation}
For the ecliptic-plane direction \(\phi=\pi/3\), one obtains
\begin{equation}
\frac{
C_X^{(0)}(\pi/2,\pi/3)
}{
C_{X,\rm stat}^{(0)}
}
=
\frac{9265}{8192}
\simeq
1.131.
\label{eq:app_ecliptic_allsky_ratio_explicit}
\end{equation}

\begin{comment}
The quantities in
Eqs.~\eqref{eq:app_pole_allsky_ratio_explicit} and
\eqref{eq:app_ecliptic_allsky_ratio_explicit} are response-coefficient
ratios. They are therefore not directly equal to the
amplitude-sensitivity ratios plotted in
Fig.~\ref{fig:sensitivity_ratio_allsky_pole_ecliptic}.    
\end{comment}

\bibliography{refs}{}

%apsrev4-2.bst 2019-01-14 (MD) hand-edited version of apsrev4-1.bst
%Control: key (0)
%Control: author (72) initials jnrlst
%Control: editor formatted (1) identically to author
%Control: production of article title (-1) disabled
%Control: page (0) single
%Control: year (1) truncated
%Control: production of eprint (0) enabled
\begin{thebibliography}{46}%
\makeatletter
\providecommand \@ifxundefined [1]{%
 \@ifx{#1\undefined}
}%
\providecommand \@ifnum [1]{%
 \ifnum #1\expandafter \@firstoftwo
 \else \expandafter \@secondoftwo
 \fi
}%
\providecommand \@ifx [1]{%
 \ifx #1\expandafter \@firstoftwo
 \else \expandafter \@secondoftwo
 \fi
}%
\providecommand \natexlab [1]{#1}%
\providecommand \enquote  [1]{``#1''}%
\providecommand \bibnamefont  [1]{#1}%
\providecommand \bibfnamefont [1]{#1}%
\providecommand \citenamefont [1]{#1}%
\providecommand \href@noop [0]{\@secondoftwo}%
\providecommand \href [0]{\begingroup \@sanitize@url \@href}%
\providecommand \@href[1]{\@@startlink{#1}\@@href}%
\providecommand \@@href[1]{\endgroup#1\@@endlink}%
\providecommand \@sanitize@url [0]{\catcode `\\12\catcode `\$12\catcode
  `\&12\catcode `\#12\catcode `\^12\catcode `\_12\catcode `\%12\relax}%
\providecommand \@@startlink[1]{}%
\providecommand \@@endlink[0]{}%
\providecommand \url  [0]{\begingroup\@sanitize@url \@url }%
\providecommand \@url [1]{\endgroup\@href {#1}{\urlprefix }}%
\providecommand \urlprefix  [0]{URL }%
\providecommand \Eprint [0]{\href }%
\providecommand \doibase [0]{https://doi.org/}%
\providecommand \selectlanguage [0]{\@gobble}%
\providecommand \bibinfo  [0]{\@secondoftwo}%
\providecommand \bibfield  [0]{\@secondoftwo}%
\providecommand \translation [1]{[#1]}%
\providecommand \BibitemOpen [0]{}%
\providecommand \bibitemStop [0]{}%
\providecommand \bibitemNoStop [0]{.\EOS\space}%
\providecommand \EOS [0]{\spacefactor3000\relax}%
\providecommand \BibitemShut  [1]{\csname bibitem#1\endcsname}%
\let\auto@bib@innerbib\@empty
%</preamble>
\bibitem [{\citenamefont {Amaro-Seoane}\ \emph {et~al.}(2017)\citenamefont
  {Amaro-Seoane} \emph {et~al.}}]{LISA}%
  \BibitemOpen
  \bibfield  {author} {\bibinfo {author} {\bibfnamefont {P.}~\bibnamefont
  {Amaro-Seoane}} \emph {et~al.},\ }\href@noop {} {\bibinfo {title} {Laser
  interferometer space antenna}} (\bibinfo {year} {2017}),\ \Eprint
  {https://arxiv.org/abs/1702.00786} {arXiv:1702.00786 [astro-ph.IM]}
  \BibitemShut {NoStop}%
\bibitem [{\citenamefont {Hu}\ and\ \citenamefont {Wu}(2017)}]{HuWu2017Taiji}%
  \BibitemOpen
  \bibfield  {author} {\bibinfo {author} {\bibfnamefont {W.-R.}\ \bibnamefont
  {Hu}}\ and\ \bibinfo {author} {\bibfnamefont {Y.-L.}\ \bibnamefont {Wu}},\
  }\href {https://doi.org/10.1093/nsr/nwx116} {\bibfield  {journal} {\bibinfo
  {journal} {Natl. Sci. Rev.}\ }\textbf {\bibinfo {volume} {4}},\ \bibinfo
  {pages} {685} (\bibinfo {year} {2017})}\BibitemShut {NoStop}%
\bibitem [{\citenamefont {Luo}\ \emph {et~al.}(2016)\citenamefont {Luo} \emph
  {et~al.}}]{TianQin}%
  \BibitemOpen
  \bibfield  {author} {\bibinfo {author} {\bibfnamefont {J.}~\bibnamefont
  {Luo}} \emph {et~al.},\ }\href
  {https://doi.org/10.1088/0264-9381/33/3/035010} {\bibfield  {journal}
  {\bibinfo  {journal} {Classical and Quantum Gravity}\ }\textbf {\bibinfo
  {volume} {33}},\ \bibinfo {pages} {035010} (\bibinfo {year}
  {2016})}\BibitemShut {NoStop}%
\bibitem [{\citenamefont {Cutler}(1998)}]{Cutler1998AngularResolution}%
  \BibitemOpen
  \bibfield  {author} {\bibinfo {author} {\bibfnamefont {C.}~\bibnamefont
  {Cutler}},\ }\href {https://doi.org/10.1103/PhysRevD.57.7089} {\bibfield
  {journal} {\bibinfo  {journal} {Phys. Rev. D}\ }\textbf {\bibinfo {volume}
  {57}},\ \bibinfo {pages} {7089} (\bibinfo {year} {1998})},\ \Eprint
  {https://arxiv.org/abs/gr-qc/9703068} {arXiv:gr-qc/9703068} \BibitemShut
  {NoStop}%
\bibitem [{\citenamefont {Cornish}\ and\ \citenamefont
  {Rubbo}(2003)}]{CornishRubbo2003}%
  \BibitemOpen
  \bibfield  {author} {\bibinfo {author} {\bibfnamefont {N.~J.}\ \bibnamefont
  {Cornish}}\ and\ \bibinfo {author} {\bibfnamefont {L.~J.}\ \bibnamefont
  {Rubbo}},\ }\href {https://doi.org/10.1103/PhysRevD.67.022001} {\bibfield
  {journal} {\bibinfo  {journal} {Phys. Rev. D}\ }\textbf {\bibinfo {volume}
  {67}},\ \bibinfo {pages} {022001} (\bibinfo {year} {2003})},\ \Eprint
  {https://arxiv.org/abs/gr-qc/0209011} {arXiv:gr-qc/0209011} \BibitemShut
  {NoStop}%
\bibitem [{\citenamefont {Tinto}\ and\ \citenamefont
  {Armstrong}(1999)}]{Tinto:1999yr}%
  \BibitemOpen
  \bibfield  {author} {\bibinfo {author} {\bibfnamefont {M.}~\bibnamefont
  {Tinto}}\ and\ \bibinfo {author} {\bibfnamefont {J.~W.}\ \bibnamefont
  {Armstrong}},\ }\href {https://doi.org/10.1103/PhysRevD.59.102003} {\bibfield
   {journal} {\bibinfo  {journal} {Phys. Rev. D}\ }\textbf {\bibinfo {volume}
  {59}},\ \bibinfo {pages} {102003} (\bibinfo {year} {1999})}\BibitemShut
  {NoStop}%
\bibitem [{\citenamefont {Petiteau}\ \emph {et~al.}(2008)\citenamefont
  {Petiteau}, \citenamefont {Auger}, \citenamefont {Halloin}, \citenamefont
  {Jeannin}, \citenamefont {Plagnol}, \citenamefont {Pireaux}, \citenamefont
  {Regimbau},\ and\ \citenamefont {Vinet}}]{Petiteau2008LISACode}%
  \BibitemOpen
  \bibfield  {author} {\bibinfo {author} {\bibfnamefont {A.}~\bibnamefont
  {Petiteau}}, \bibinfo {author} {\bibfnamefont {G.}~\bibnamefont {Auger}},
  \bibinfo {author} {\bibfnamefont {H.}~\bibnamefont {Halloin}}, \bibinfo
  {author} {\bibfnamefont {O.}~\bibnamefont {Jeannin}}, \bibinfo {author}
  {\bibfnamefont {E.}~\bibnamefont {Plagnol}}, \bibinfo {author} {\bibfnamefont
  {S.}~\bibnamefont {Pireaux}}, \bibinfo {author} {\bibfnamefont
  {T.}~\bibnamefont {Regimbau}},\ and\ \bibinfo {author} {\bibfnamefont
  {J.-Y.}\ \bibnamefont {Vinet}},\ }\href
  {https://doi.org/10.1103/PhysRevD.77.023002} {\bibfield  {journal} {\bibinfo
  {journal} {Phys. Rev. D}\ }\textbf {\bibinfo {volume} {77}},\ \bibinfo
  {pages} {023002} (\bibinfo {year} {2008})},\ \Eprint
  {https://arxiv.org/abs/0802.2023} {arXiv:0802.2023 [gr-qc]} \BibitemShut
  {NoStop}%
\bibitem [{\citenamefont {Larson}\ \emph {et~al.}(2000)\citenamefont {Larson},
  \citenamefont {Hiscock},\ and\ \citenamefont
  {Hellings}}]{Larson2000Sensitivity}%
  \BibitemOpen
  \bibfield  {author} {\bibinfo {author} {\bibfnamefont {S.~L.}\ \bibnamefont
  {Larson}}, \bibinfo {author} {\bibfnamefont {W.~A.}\ \bibnamefont
  {Hiscock}},\ and\ \bibinfo {author} {\bibfnamefont {R.~W.}\ \bibnamefont
  {Hellings}},\ }\href {https://doi.org/10.1103/PhysRevD.62.062001} {\bibfield
  {journal} {\bibinfo  {journal} {Phys. Rev. D}\ }\textbf {\bibinfo {volume}
  {62}},\ \bibinfo {pages} {062001} (\bibinfo {year} {2000})},\ \Eprint
  {https://arxiv.org/abs/gr-qc/9909080} {arXiv:gr-qc/9909080} \BibitemShut
  {NoStop}%
\bibitem [{\citenamefont {Robson}\ \emph {et~al.}(2019)\citenamefont {Robson},
  \citenamefont {Cornish},\ and\ \citenamefont {Liu}}]{Robson:2018ifk}%
  \BibitemOpen
  \bibfield  {author} {\bibinfo {author} {\bibfnamefont {T.}~\bibnamefont
  {Robson}}, \bibinfo {author} {\bibfnamefont {N.~J.}\ \bibnamefont
  {Cornish}},\ and\ \bibinfo {author} {\bibfnamefont {C.}~\bibnamefont {Liu}},\
  }\href {https://doi.org/10.1088/1361-6382/ab1101} {\bibfield  {journal}
  {\bibinfo  {journal} {Class. Quant. Grav.}\ }\textbf {\bibinfo {volume}
  {36}},\ \bibinfo {pages} {105011} (\bibinfo {year} {2019})},\ \Eprint
  {https://arxiv.org/abs/1803.01944} {arXiv:1803.01944 [astro-ph.HE]}
  \BibitemShut {NoStop}%
\bibitem [{\citenamefont {Smith}\ and\ \citenamefont
  {Caldwell}(2019)}]{Smith:2019wny}%
  \BibitemOpen
  \bibfield  {author} {\bibinfo {author} {\bibfnamefont {T.~L.}\ \bibnamefont
  {Smith}}\ and\ \bibinfo {author} {\bibfnamefont {R.}~\bibnamefont
  {Caldwell}},\ }\href {https://doi.org/10.1103/PhysRevD.100.104055} {\bibfield
   {journal} {\bibinfo  {journal} {Phys. Rev. D}\ }\textbf {\bibinfo {volume}
  {100}},\ \bibinfo {pages} {104055} (\bibinfo {year} {2019})},\ \bibinfo
  {note} {[Erratum: Phys.Rev.D 105, 029902 (2022)]},\ \Eprint
  {https://arxiv.org/abs/1908.00546} {arXiv:1908.00546 [astro-ph.CO]}
  \BibitemShut {NoStop}%
\bibitem [{\citenamefont {Zhang}\ \emph {et~al.}(2019)\citenamefont {Zhang},
  \citenamefont {Gao}, \citenamefont {Gong}, \citenamefont {Liang},
  \citenamefont {Weinstein},\ and\ \citenamefont {Zhang}}]{Zhang:2019oet}%
  \BibitemOpen
  \bibfield  {author} {\bibinfo {author} {\bibfnamefont {C.}~\bibnamefont
  {Zhang}}, \bibinfo {author} {\bibfnamefont {Q.}~\bibnamefont {Gao}}, \bibinfo
  {author} {\bibfnamefont {Y.}~\bibnamefont {Gong}}, \bibinfo {author}
  {\bibfnamefont {D.}~\bibnamefont {Liang}}, \bibinfo {author} {\bibfnamefont
  {A.~J.}\ \bibnamefont {Weinstein}},\ and\ \bibinfo {author} {\bibfnamefont
  {C.}~\bibnamefont {Zhang}},\ }\href
  {https://doi.org/10.1103/PhysRevD.100.064033} {\bibfield  {journal} {\bibinfo
   {journal} {Phys. Rev. D}\ }\textbf {\bibinfo {volume} {100}},\ \bibinfo
  {pages} {064033} (\bibinfo {year} {2019})},\ \Eprint
  {https://arxiv.org/abs/1906.10901} {arXiv:1906.10901 [gr-qc]} \BibitemShut
  {NoStop}%
\bibitem [{\citenamefont {Liang}\ \emph {et~al.}(2019)\citenamefont {Liang},
  \citenamefont {Gong}, \citenamefont {Weinstein}, \citenamefont {Zhang},\ and\
  \citenamefont {Zhang}}]{Liang:2019pry}%
  \BibitemOpen
  \bibfield  {author} {\bibinfo {author} {\bibfnamefont {D.}~\bibnamefont
  {Liang}}, \bibinfo {author} {\bibfnamefont {Y.}~\bibnamefont {Gong}},
  \bibinfo {author} {\bibfnamefont {A.~J.}\ \bibnamefont {Weinstein}}, \bibinfo
  {author} {\bibfnamefont {C.}~\bibnamefont {Zhang}},\ and\ \bibinfo {author}
  {\bibfnamefont {C.}~\bibnamefont {Zhang}},\ }\href
  {https://doi.org/10.1103/PhysRevD.99.104027} {\bibfield  {journal} {\bibinfo
  {journal} {Phys. Rev. D}\ }\textbf {\bibinfo {volume} {99}},\ \bibinfo
  {pages} {104027} (\bibinfo {year} {2019})},\ \Eprint
  {https://arxiv.org/abs/1901.09624} {arXiv:1901.09624 [gr-qc]} \BibitemShut
  {NoStop}%
\bibitem [{\citenamefont {Zhang}\ \emph {et~al.}(2020)\citenamefont {Zhang},
  \citenamefont {Gao}, \citenamefont {Gong}, \citenamefont {Wang},
  \citenamefont {Weinstein},\ and\ \citenamefont {Zhang}}]{Zhang:2020khm}%
  \BibitemOpen
  \bibfield  {author} {\bibinfo {author} {\bibfnamefont {C.}~\bibnamefont
  {Zhang}}, \bibinfo {author} {\bibfnamefont {Q.}~\bibnamefont {Gao}}, \bibinfo
  {author} {\bibfnamefont {Y.}~\bibnamefont {Gong}}, \bibinfo {author}
  {\bibfnamefont {B.}~\bibnamefont {Wang}}, \bibinfo {author} {\bibfnamefont
  {A.~J.}\ \bibnamefont {Weinstein}},\ and\ \bibinfo {author} {\bibfnamefont
  {C.}~\bibnamefont {Zhang}},\ }\href
  {https://doi.org/10.1103/PhysRevD.101.124027} {\bibfield  {journal} {\bibinfo
   {journal} {Phys. Rev. D}\ }\textbf {\bibinfo {volume} {101}},\ \bibinfo
  {pages} {124027} (\bibinfo {year} {2020})},\ \Eprint
  {https://arxiv.org/abs/2003.01441} {arXiv:2003.01441 [gr-qc]} \BibitemShut
  {NoStop}%
\bibitem [{\citenamefont {Lu}\ \emph {et~al.}(2019)\citenamefont {Lu},
  \citenamefont {Tan},\ and\ \citenamefont {Shao}}]{Lu:2019log}%
  \BibitemOpen
  \bibfield  {author} {\bibinfo {author} {\bibfnamefont {X.-Y.}\ \bibnamefont
  {Lu}}, \bibinfo {author} {\bibfnamefont {Y.-J.}\ \bibnamefont {Tan}},\ and\
  \bibinfo {author} {\bibfnamefont {C.-G.}\ \bibnamefont {Shao}},\ }\href
  {https://doi.org/10.1103/PhysRevD.100.044042} {\bibfield  {journal} {\bibinfo
   {journal} {Phys. Rev. D}\ }\textbf {\bibinfo {volume} {100}},\ \bibinfo
  {pages} {044042} (\bibinfo {year} {2019})},\ \Eprint
  {https://arxiv.org/abs/2007.03400} {arXiv:2007.03400 [gr-qc]} \BibitemShut
  {NoStop}%
\bibitem [{\citenamefont {Babak}\ \emph {et~al.}(2021)\citenamefont {Babak},
  \citenamefont {Petiteau},\ and\ \citenamefont {Hewitson}}]{Babak:2021mhe}%
  \BibitemOpen
  \bibfield  {author} {\bibinfo {author} {\bibfnamefont {S.}~\bibnamefont
  {Babak}}, \bibinfo {author} {\bibfnamefont {A.}~\bibnamefont {Petiteau}},\
  and\ \bibinfo {author} {\bibfnamefont {M.}~\bibnamefont {Hewitson}},\
  }\href@noop {} {\bibfield  {journal} {\bibinfo  {journal} {arXiv preprint}\ }
  (\bibinfo {year} {2021})},\ \Eprint {https://arxiv.org/abs/2108.01167}
  {arXiv:2108.01167 [astro-ph.IM]} \BibitemShut {NoStop}%
\bibitem [{\citenamefont {Wang}\ \emph
  {et~al.}(2021{\natexlab{a}})\citenamefont {Wang}, \citenamefont {Tan},
  \citenamefont {Qian},\ and\ \citenamefont {Shao}}]{Wang:2021jsv}%
  \BibitemOpen
  \bibfield  {author} {\bibinfo {author} {\bibfnamefont {P.-P.}\ \bibnamefont
  {Wang}}, \bibinfo {author} {\bibfnamefont {Y.-J.}\ \bibnamefont {Tan}},
  \bibinfo {author} {\bibfnamefont {W.-L.}\ \bibnamefont {Qian}},\ and\
  \bibinfo {author} {\bibfnamefont {C.-G.}\ \bibnamefont {Shao}},\ }\href
  {https://doi.org/10.1103/PhysRevD.103.063021} {\bibfield  {journal} {\bibinfo
   {journal} {Phys. Rev. D}\ }\textbf {\bibinfo {volume} {103}},\ \bibinfo
  {pages} {063021} (\bibinfo {year} {2021}{\natexlab{a}})}\BibitemShut
  {NoStop}%
\bibitem [{\citenamefont {Wang}\ \emph {et~al.}(2022)\citenamefont {Wang},
  \citenamefont {Li}, \citenamefont {Xu},\ and\ \citenamefont
  {Fan}}]{Wang:2022sti}%
  \BibitemOpen
  \bibfield  {author} {\bibinfo {author} {\bibfnamefont {G.}~\bibnamefont
  {Wang}}, \bibinfo {author} {\bibfnamefont {B.}~\bibnamefont {Li}}, \bibinfo
  {author} {\bibfnamefont {P.}~\bibnamefont {Xu}},\ and\ \bibinfo {author}
  {\bibfnamefont {X.}~\bibnamefont {Fan}},\ }\href
  {https://doi.org/10.1103/PhysRevD.106.044054} {\bibfield  {journal} {\bibinfo
   {journal} {Phys. Rev. D}\ }\textbf {\bibinfo {volume} {106}},\ \bibinfo
  {pages} {044054} (\bibinfo {year} {2022})},\ \Eprint
  {https://arxiv.org/abs/2201.10902} {arXiv:2201.10902 [gr-qc]} \BibitemShut
  {NoStop}%
\bibitem [{\citenamefont {Du}\ \emph {et~al.}(2026)\citenamefont {Du},
  \citenamefont {Wang}, \citenamefont {Luo}, \citenamefont {Han}, \citenamefont
  {Zhang} \emph {et~al.}}]{Du2026TaijiPipeline}%
  \BibitemOpen
  \bibfield  {author} {\bibinfo {author} {\bibfnamefont {M.}~\bibnamefont
  {Du}}, \bibinfo {author} {\bibfnamefont {P.}~\bibnamefont {Wang}}, \bibinfo
  {author} {\bibfnamefont {Z.}~\bibnamefont {Luo}}, \bibinfo {author}
  {\bibfnamefont {W.-B.}\ \bibnamefont {Han}}, \bibinfo {author} {\bibfnamefont
  {X.}~\bibnamefont {Zhang}}, \emph {et~al.},\ }\href
  {https://doi.org/10.1007/s11433-025-2870-8} {\bibfield  {journal} {\bibinfo
  {journal} {Sci. China Phys. Mech. Astron.}\ }\textbf {\bibinfo {volume}
  {69}},\ \bibinfo {pages} {249501} (\bibinfo {year} {2026})},\ \Eprint
  {https://arxiv.org/abs/2505.16500} {arXiv:2505.16500 [gr-qc]} \BibitemShut
  {NoStop}%
\bibitem [{\citenamefont {Adams}\ and\ \citenamefont
  {Cornish}(2010)}]{Adams:2010vc}%
  \BibitemOpen
  \bibfield  {author} {\bibinfo {author} {\bibfnamefont {M.~R.}\ \bibnamefont
  {Adams}}\ and\ \bibinfo {author} {\bibfnamefont {N.~J.}\ \bibnamefont
  {Cornish}},\ }\href {https://doi.org/10.1103/PhysRevD.82.022002} {\bibfield
  {journal} {\bibinfo  {journal} {Phys. Rev. D}\ }\textbf {\bibinfo {volume}
  {82}},\ \bibinfo {pages} {022002} (\bibinfo {year} {2010})},\ \Eprint
  {https://arxiv.org/abs/1002.1291} {arXiv:1002.1291 [gr-qc]} \BibitemShut
  {NoStop}%
\bibitem [{\citenamefont {Wang}\ and\ \citenamefont {Ni}(2023)}]{Wang:2020fwa}%
  \BibitemOpen
  \bibfield  {author} {\bibinfo {author} {\bibfnamefont {G.}~\bibnamefont
  {Wang}}\ and\ \bibinfo {author} {\bibfnamefont {W.-T.}\ \bibnamefont {Ni}},\
  }\href {https://doi.org/10.1088/1402-4896/acd882} {\bibfield  {journal}
  {\bibinfo  {journal} {Phys. Scripta}\ }\textbf {\bibinfo {volume} {98}},\
  \bibinfo {pages} {075005} (\bibinfo {year} {2023})},\ \Eprint
  {https://arxiv.org/abs/2008.05812} {arXiv:2008.05812 [gr-qc]} \BibitemShut
  {NoStop}%
\bibitem [{\citenamefont {Wang}\ \emph
  {et~al.}(2021{\natexlab{b}})\citenamefont {Wang}, \citenamefont {Ni},
  \citenamefont {Han},\ and\ \citenamefont {Qiao}}]{Wang:2020pkk}%
  \BibitemOpen
  \bibfield  {author} {\bibinfo {author} {\bibfnamefont {G.}~\bibnamefont
  {Wang}}, \bibinfo {author} {\bibfnamefont {W.-T.}\ \bibnamefont {Ni}},
  \bibinfo {author} {\bibfnamefont {W.-B.}\ \bibnamefont {Han}},\ and\ \bibinfo
  {author} {\bibfnamefont {C.-F.}\ \bibnamefont {Qiao}},\ }\href
  {https://doi.org/10.1103/PhysRevD.103.122006} {\bibfield  {journal} {\bibinfo
   {journal} {Phys. Rev. D}\ }\textbf {\bibinfo {volume} {103}},\ \bibinfo
  {pages} {122006} (\bibinfo {year} {2021}{\natexlab{b}})},\ \Eprint
  {https://arxiv.org/abs/2010.15544} {arXiv:2010.15544 [gr-qc]} \BibitemShut
  {NoStop}%
\bibitem [{\citenamefont {Tinto}\ and\ \citenamefont
  {Dhurandhar}(2005)}]{TintoDhurandhar2005TDI}%
  \BibitemOpen
  \bibfield  {author} {\bibinfo {author} {\bibfnamefont {M.}~\bibnamefont
  {Tinto}}\ and\ \bibinfo {author} {\bibfnamefont {S.~V.}\ \bibnamefont
  {Dhurandhar}},\ }\href {https://doi.org/10.12942/lrr-2005-4} {\bibfield
  {journal} {\bibinfo  {journal} {Living Rev. Relativity}\ }\textbf {\bibinfo
  {volume} {8}},\ \bibinfo {pages} {4} (\bibinfo {year} {2005})},\ \Eprint
  {https://arxiv.org/abs/gr-qc/0409034} {arXiv:gr-qc/0409034} \BibitemShut
  {NoStop}%
\bibitem [{\citenamefont {Estabrook}\ and\ \citenamefont
  {Wahlquist}(1975)}]{EstabrookWahlquist1975}%
  \BibitemOpen
  \bibfield  {author} {\bibinfo {author} {\bibfnamefont {F.~B.}\ \bibnamefont
  {Estabrook}}\ and\ \bibinfo {author} {\bibfnamefont {H.~D.}\ \bibnamefont
  {Wahlquist}},\ }\href {https://doi.org/10.1007/BF00762449} {\bibfield
  {journal} {\bibinfo  {journal} {General Relativity and Gravitation}\ }\textbf
  {\bibinfo {volume} {6}},\ \bibinfo {pages} {439} (\bibinfo {year}
  {1975})}\BibitemShut {NoStop}%
\bibitem [{\citenamefont {Vallisneri}(2005{\natexlab{a}})}]{Vallisneri2005}%
  \BibitemOpen
  \bibfield  {author} {\bibinfo {author} {\bibfnamefont {M.}~\bibnamefont
  {Vallisneri}},\ }\href {https://doi.org/10.1103/PhysRevD.72.042003}
  {\bibfield  {journal} {\bibinfo  {journal} {Physical Review D}\ }\textbf
  {\bibinfo {volume} {72}},\ \bibinfo {pages} {042003} (\bibinfo {year}
  {2005}{\natexlab{a}})},\ \Eprint {https://arxiv.org/abs/gr-qc/0504145}
  {arXiv:gr-qc/0504145 [gr-qc]} \BibitemShut {NoStop}%
\bibitem [{\citenamefont {Rubbo}\ \emph {et~al.}(2004)\citenamefont {Rubbo},
  \citenamefont {Cornish},\ and\ \citenamefont
  {Poujade}}]{RubboCornishPoujade2004}%
  \BibitemOpen
  \bibfield  {author} {\bibinfo {author} {\bibfnamefont {L.~J.}\ \bibnamefont
  {Rubbo}}, \bibinfo {author} {\bibfnamefont {N.~J.}\ \bibnamefont {Cornish}},\
  and\ \bibinfo {author} {\bibfnamefont {O.}~\bibnamefont {Poujade}},\ }\href
  {https://doi.org/10.1103/PhysRevD.69.082003} {\bibfield  {journal} {\bibinfo
  {journal} {Physical Review D}\ }\textbf {\bibinfo {volume} {69}},\ \bibinfo
  {pages} {082003} (\bibinfo {year} {2004})},\ \Eprint
  {https://arxiv.org/abs/gr-qc/0311069} {arXiv:gr-qc/0311069 [gr-qc]}
  \BibitemShut {NoStop}%
\bibitem [{\citenamefont {Tinto}\ \emph {et~al.}(2002)\citenamefont {Tinto},
  \citenamefont {Estabrook},\ and\ \citenamefont {Armstrong}}]{Tinto:2002de}%
  \BibitemOpen
  \bibfield  {author} {\bibinfo {author} {\bibfnamefont {M.}~\bibnamefont
  {Tinto}}, \bibinfo {author} {\bibfnamefont {F.~B.}\ \bibnamefont
  {Estabrook}},\ and\ \bibinfo {author} {\bibfnamefont {J.~W.}\ \bibnamefont
  {Armstrong}},\ }\href {https://doi.org/10.1103/PhysRevD.65.082003} {\bibfield
   {journal} {\bibinfo  {journal} {Phys. Rev. D}\ }\textbf {\bibinfo {volume}
  {65}},\ \bibinfo {pages} {082003} (\bibinfo {year} {2002})}\BibitemShut
  {NoStop}%
\bibitem [{\citenamefont {Vallisneri}(2005{\natexlab{b}})}]{Vallisneri:2005ji}%
  \BibitemOpen
  \bibfield  {author} {\bibinfo {author} {\bibfnamefont {M.}~\bibnamefont
  {Vallisneri}},\ }\href {https://doi.org/10.1103/PhysRevD.76.109903}
  {\bibfield  {journal} {\bibinfo  {journal} {Phys. Rev. D}\ }\textbf {\bibinfo
  {volume} {72}},\ \bibinfo {pages} {042003} (\bibinfo {year}
  {2005}{\natexlab{b}})},\ \bibinfo {note} {[Erratum: Phys.Rev.D 76, 109903
  (2007)]},\ \Eprint {https://arxiv.org/abs/gr-qc/0504145}
  {arXiv:gr-qc/0504145} \BibitemShut {NoStop}%
\bibitem [{\citenamefont {Tinto}\ \emph {et~al.}(2023)\citenamefont {Tinto},
  \citenamefont {Dhurandhar},\ and\ \citenamefont {Malakar}}]{Tinto:2022zmf}%
  \BibitemOpen
  \bibfield  {author} {\bibinfo {author} {\bibfnamefont {M.}~\bibnamefont
  {Tinto}}, \bibinfo {author} {\bibfnamefont {S.}~\bibnamefont {Dhurandhar}},\
  and\ \bibinfo {author} {\bibfnamefont {D.}~\bibnamefont {Malakar}},\ }\href
  {https://doi.org/10.1103/PhysRevD.107.082001} {\bibfield  {journal} {\bibinfo
   {journal} {Phys. Rev. D}\ }\textbf {\bibinfo {volume} {107}},\ \bibinfo
  {pages} {082001} (\bibinfo {year} {2023})},\ \Eprint
  {https://arxiv.org/abs/2212.05967} {arXiv:2212.05967 [gr-qc]} \BibitemShut
  {NoStop}%
\bibitem [{\citenamefont {Pierce}\ \emph {et~al.}(2018)\citenamefont {Pierce},
  \citenamefont {Riles},\ and\ \citenamefont {Zhao}}]{Pierce:2018xmy}%
  \BibitemOpen
  \bibfield  {author} {\bibinfo {author} {\bibfnamefont {A.}~\bibnamefont
  {Pierce}}, \bibinfo {author} {\bibfnamefont {K.}~\bibnamefont {Riles}},\ and\
  \bibinfo {author} {\bibfnamefont {Y.}~\bibnamefont {Zhao}},\ }\href
  {https://doi.org/10.1103/PhysRevLett.121.061102} {\bibfield  {journal}
  {\bibinfo  {journal} {Phys. Rev. Lett.}\ }\textbf {\bibinfo {volume} {121}},\
  \bibinfo {pages} {061102} (\bibinfo {year} {2018})},\ \Eprint
  {https://arxiv.org/abs/1801.10161} {arXiv:1801.10161 [hep-ph]} \BibitemShut
  {NoStop}%
\bibitem [{\citenamefont {Morisaki}\ and\ \citenamefont
  {Suyama}(2019)}]{Morisaki:2018htj}%
  \BibitemOpen
  \bibfield  {author} {\bibinfo {author} {\bibfnamefont {S.}~\bibnamefont
  {Morisaki}}\ and\ \bibinfo {author} {\bibfnamefont {T.}~\bibnamefont
  {Suyama}},\ }\href {https://doi.org/10.1103/PhysRevD.100.123512} {\bibfield
  {journal} {\bibinfo  {journal} {Phys. Rev. D}\ }\textbf {\bibinfo {volume}
  {100}},\ \bibinfo {pages} {123512} (\bibinfo {year} {2019})},\ \Eprint
  {https://arxiv.org/abs/1811.05003} {arXiv:1811.05003 [hep-ph]} \BibitemShut
  {NoStop}%
\bibitem [{\citenamefont {Grote}\ and\ \citenamefont
  {Stadnik}(2019)}]{Grote:2019uvn}%
  \BibitemOpen
  \bibfield  {author} {\bibinfo {author} {\bibfnamefont {H.}~\bibnamefont
  {Grote}}\ and\ \bibinfo {author} {\bibfnamefont {Y.~V.}\ \bibnamefont
  {Stadnik}},\ }\href {https://doi.org/10.1103/PhysRevResearch.1.033187}
  {\bibfield  {journal} {\bibinfo  {journal} {Phys. Rev. Res.}\ }\textbf
  {\bibinfo {volume} {1}},\ \bibinfo {pages} {033187} (\bibinfo {year}
  {2019})},\ \Eprint {https://arxiv.org/abs/1906.06193} {arXiv:1906.06193
  [astro-ph.IM]} \BibitemShut {NoStop}%
\bibitem [{\citenamefont {Miller}\ and\ \citenamefont
  {Mendes}(2023)}]{Miller:2023kkd}%
  \BibitemOpen
  \bibfield  {author} {\bibinfo {author} {\bibfnamefont {A.~L.}\ \bibnamefont
  {Miller}}\ and\ \bibinfo {author} {\bibfnamefont {L.}~\bibnamefont
  {Mendes}},\ }\href {https://doi.org/10.1103/PhysRevD.107.063015} {\bibfield
  {journal} {\bibinfo  {journal} {Phys. Rev. D}\ }\textbf {\bibinfo {volume}
  {107}},\ \bibinfo {pages} {063015} (\bibinfo {year} {2023})},\ \Eprint
  {https://arxiv.org/abs/2301.08736} {arXiv:2301.08736 [gr-qc]} \BibitemShut
  {NoStop}%
\bibitem [{\citenamefont {Yu}\ \emph {et~al.}(2023)\citenamefont {Yu},
  \citenamefont {Yao}, \citenamefont {Tang},\ and\ \citenamefont
  {Wu}}]{Yu:2023iog}%
  \BibitemOpen
  \bibfield  {author} {\bibinfo {author} {\bibfnamefont {J.-C.}\ \bibnamefont
  {Yu}}, \bibinfo {author} {\bibfnamefont {Y.-H.}\ \bibnamefont {Yao}},
  \bibinfo {author} {\bibfnamefont {Y.}~\bibnamefont {Tang}},\ and\ \bibinfo
  {author} {\bibfnamefont {Y.-L.}\ \bibnamefont {Wu}},\ }\href
  {https://doi.org/10.1103/PhysRevD.108.083007} {\bibfield  {journal} {\bibinfo
   {journal} {Phys. Rev. D}\ }\textbf {\bibinfo {volume} {108}},\ \bibinfo
  {pages} {083007} (\bibinfo {year} {2023})},\ \Eprint
  {https://arxiv.org/abs/2307.09197} {arXiv:2307.09197 [gr-qc]} \BibitemShut
  {NoStop}%
\bibitem [{\citenamefont {Yao}\ and\ \citenamefont {Tang}(2024)}]{Yao:2024fie}%
  \BibitemOpen
  \bibfield  {author} {\bibinfo {author} {\bibfnamefont {Y.-H.}\ \bibnamefont
  {Yao}}\ and\ \bibinfo {author} {\bibfnamefont {Y.}~\bibnamefont {Tang}},\
  }\href {https://doi.org/10.1103/PhysRevD.110.095015} {\bibfield  {journal}
  {\bibinfo  {journal} {Phys. Rev. D}\ }\textbf {\bibinfo {volume} {110}},\
  \bibinfo {pages} {095015} (\bibinfo {year} {2024})},\ \Eprint
  {https://arxiv.org/abs/2404.01494} {arXiv:2404.01494 [hep-ph]} \BibitemShut
  {NoStop}%
\bibitem [{\citenamefont {Yao}\ \emph {et~al.}(2025{\natexlab{a}})\citenamefont
  {Yao}, \citenamefont {Jiang},\ and\ \citenamefont {Tang}}]{Yao:2024hap}%
  \BibitemOpen
  \bibfield  {author} {\bibinfo {author} {\bibfnamefont {Y.-H.}\ \bibnamefont
  {Yao}}, \bibinfo {author} {\bibfnamefont {T.}~\bibnamefont {Jiang}},\ and\
  \bibinfo {author} {\bibfnamefont {Y.}~\bibnamefont {Tang}},\ }\href
  {https://doi.org/10.1103/PhysRevD.111.055031} {\bibfield  {journal} {\bibinfo
   {journal} {Phys. Rev. D}\ }\textbf {\bibinfo {volume} {111}},\ \bibinfo
  {pages} {055031} (\bibinfo {year} {2025}{\natexlab{a}})},\ \Eprint
  {https://arxiv.org/abs/2410.22072} {arXiv:2410.22072 [hep-ph]} \BibitemShut
  {NoStop}%
\bibitem [{\citenamefont {Xu}\ \emph {et~al.}(2025)\citenamefont {Xu},
  \citenamefont {Yao}, \citenamefont {Tang},\ and\ \citenamefont
  {Wu}}]{Xu:2025rfv}%
  \BibitemOpen
  \bibfield  {author} {\bibinfo {author} {\bibfnamefont {H.-T.}\ \bibnamefont
  {Xu}}, \bibinfo {author} {\bibfnamefont {Y.-H.}\ \bibnamefont {Yao}},
  \bibinfo {author} {\bibfnamefont {Y.}~\bibnamefont {Tang}},\ and\ \bibinfo
  {author} {\bibfnamefont {Y.-L.}\ \bibnamefont {Wu}},\ }\href
  {https://doi.org/10.1103/xlqy-r6n3} {\bibfield  {journal} {\bibinfo
  {journal} {Phys. Rev. D}\ }\textbf {\bibinfo {volume} {112}},\ \bibinfo
  {pages} {095021} (\bibinfo {year} {2025})},\ \Eprint
  {https://arxiv.org/abs/2506.09744} {arXiv:2506.09744 [hep-ph]} \BibitemShut
  {NoStop}%
\bibitem [{\citenamefont {Gu{\'e}}\ \emph {et~al.}(2025)\citenamefont
  {Gu{\'e}}, \citenamefont {Hees},\ and\ \citenamefont {Wolf}}]{Gue:2024txz}%
  \BibitemOpen
  \bibfield  {author} {\bibinfo {author} {\bibfnamefont {J.}~\bibnamefont
  {Gu{\'e}}}, \bibinfo {author} {\bibfnamefont {A.}~\bibnamefont {Hees}},\ and\
  \bibinfo {author} {\bibfnamefont {P.}~\bibnamefont {Wolf}},\ }\href
  {https://doi.org/10.1088/1361-6382/adb23c} {\bibfield  {journal} {\bibinfo
  {journal} {Class. Quant. Grav.}\ }\textbf {\bibinfo {volume} {42}},\ \bibinfo
  {pages} {055015} (\bibinfo {year} {2025})},\ \Eprint
  {https://arxiv.org/abs/2410.17763} {arXiv:2410.17763 [hep-ph]} \BibitemShut
  {NoStop}%
\bibitem [{\citenamefont {Yao}\ \emph {et~al.}(2025{\natexlab{b}})\citenamefont
  {Yao}, \citenamefont {Jiang}, \citenamefont {Ren}, \citenamefont {Chen},
  \citenamefont {Tang},\ and\ \citenamefont {Zhou}}]{Yao:2025vgy}%
  \BibitemOpen
  \bibfield  {author} {\bibinfo {author} {\bibfnamefont {Y.-H.}\ \bibnamefont
  {Yao}}, \bibinfo {author} {\bibfnamefont {T.}~\bibnamefont {Jiang}}, \bibinfo
  {author} {\bibfnamefont {W.}~\bibnamefont {Ren}}, \bibinfo {author}
  {\bibfnamefont {D.}~\bibnamefont {Chen}}, \bibinfo {author} {\bibfnamefont
  {Y.}~\bibnamefont {Tang}},\ and\ \bibinfo {author} {\bibfnamefont {Y.-F.}\
  \bibnamefont {Zhou}},\ }\href@noop {} {\  (\bibinfo {year}
  {2025}{\natexlab{b}})},\ \Eprint {https://arxiv.org/abs/2508.14655}
  {arXiv:2508.14655 [hep-ph]} \BibitemShut {NoStop}%
\bibitem [{\citenamefont {Yao}\ \emph {et~al.}(2026)\citenamefont {Yao},
  \citenamefont {Bi}, \citenamefont {Yin},\ and\ \citenamefont
  {Huang}}]{Yao:2025wfd}%
  \BibitemOpen
  \bibfield  {author} {\bibinfo {author} {\bibfnamefont {R.-M.}\ \bibnamefont
  {Yao}}, \bibinfo {author} {\bibfnamefont {X.-J.}\ \bibnamefont {Bi}},
  \bibinfo {author} {\bibfnamefont {P.-F.}\ \bibnamefont {Yin}},\ and\ \bibinfo
  {author} {\bibfnamefont {Q.-G.}\ \bibnamefont {Huang}},\ }\href
  {https://doi.org/10.1088/1475-7516/2026/04/082} {\bibfield  {journal}
  {\bibinfo  {journal} {JCAP}\ }\textbf {\bibinfo {volume} {04}},\ \bibinfo
  {pages} {082}},\ \Eprint {https://arxiv.org/abs/2504.10083} {arXiv:2504.10083
  [hep-ph]} \BibitemShut {NoStop}%
\bibitem [{\citenamefont {Liu}\ \emph {et~al.}(2026)\citenamefont {Liu},
  \citenamefont {Zhang}, \citenamefont {Du}, \citenamefont {Liu}, \citenamefont
  {Xu},\ and\ \citenamefont {Zhang}}]{Liu:2025hwn}%
  \BibitemOpen
  \bibfield  {author} {\bibinfo {author} {\bibfnamefont {Y.-Y.}\ \bibnamefont
  {Liu}}, \bibinfo {author} {\bibfnamefont {J.-R.}\ \bibnamefont {Zhang}},
  \bibinfo {author} {\bibfnamefont {M.-H.}\ \bibnamefont {Du}}, \bibinfo
  {author} {\bibfnamefont {H.-S.}\ \bibnamefont {Liu}}, \bibinfo {author}
  {\bibfnamefont {P.}~\bibnamefont {Xu}},\ and\ \bibinfo {author}
  {\bibfnamefont {Y.-L.}\ \bibnamefont {Zhang}},\ }\href
  {https://doi.org/10.1140/epjc/s10052-026-15578-3} {\bibfield  {journal}
  {\bibinfo  {journal} {Eur. Phys. J. C}\ }\textbf {\bibinfo {volume} {86}},\
  \bibinfo {pages} {347} (\bibinfo {year} {2026})},\ \Eprint
  {https://arxiv.org/abs/2511.15438} {arXiv:2511.15438 [gr-qc]} \BibitemShut
  {NoStop}%
\bibitem [{\citenamefont {Zhang}\ \emph {et~al.}(2025)\citenamefont {Zhang},
  \citenamefont {Chen}, \citenamefont {Jiao}, \citenamefont {Cai},\ and\
  \citenamefont {Zhang}}]{Zhang:2025fck}%
  \BibitemOpen
  \bibfield  {author} {\bibinfo {author} {\bibfnamefont {J.-R.}\ \bibnamefont
  {Zhang}}, \bibinfo {author} {\bibfnamefont {J.}~\bibnamefont {Chen}},
  \bibinfo {author} {\bibfnamefont {H.-S.}\ \bibnamefont {Jiao}}, \bibinfo
  {author} {\bibfnamefont {R.-G.}\ \bibnamefont {Cai}},\ and\ \bibinfo {author}
  {\bibfnamefont {Y.-L.}\ \bibnamefont {Zhang}},\ }\href
  {https://doi.org/10.1103/cz2w-5cfj} {\bibfield  {journal} {\bibinfo
  {journal} {Phys. Rev. D}\ }\textbf {\bibinfo {volume} {112}},\ \bibinfo
  {pages} {064030} (\bibinfo {year} {2025})},\ \Eprint
  {https://arxiv.org/abs/2501.11071} {arXiv:2501.11071 [gr-qc]} \BibitemShut
  {NoStop}%
\bibitem [{\citenamefont {Chen}\ \emph {et~al.}(2026)\citenamefont {Chen},
  \citenamefont {Wang}, \citenamefont {Wang}, \citenamefont {Luo},\ and\
  \citenamefont {Shao}}]{Chen:2026mef}%
  \BibitemOpen
  \bibfield  {author} {\bibinfo {author} {\bibfnamefont {Y.}~\bibnamefont
  {Chen}}, \bibinfo {author} {\bibfnamefont {P.-P.}\ \bibnamefont {Wang}},
  \bibinfo {author} {\bibfnamefont {B.}~\bibnamefont {Wang}}, \bibinfo {author}
  {\bibfnamefont {R.}~\bibnamefont {Luo}},\ and\ \bibinfo {author}
  {\bibfnamefont {C.-G.}\ \bibnamefont {Shao}},\ }\href
  {https://doi.org/10.3390/universe12020048} {\bibfield  {journal} {\bibinfo
  {journal} {Universe}\ }\textbf {\bibinfo {volume} {12}},\ \bibinfo {pages}
  {48} (\bibinfo {year} {2026})},\ \Eprint {https://arxiv.org/abs/2603.07158}
  {arXiv:2603.07158 [astro-ph.CO]} \BibitemShut {NoStop}%
\bibitem [{\citenamefont {Prince}\ \emph {et~al.}(2002)\citenamefont {Prince},
  \citenamefont {Tinto}, \citenamefont {Larson},\ and\ \citenamefont
  {Armstrong}}]{Prince:2002hp}%
  \BibitemOpen
  \bibfield  {author} {\bibinfo {author} {\bibfnamefont {T.~A.}\ \bibnamefont
  {Prince}}, \bibinfo {author} {\bibfnamefont {M.}~\bibnamefont {Tinto}},
  \bibinfo {author} {\bibfnamefont {S.~L.}\ \bibnamefont {Larson}},\ and\
  \bibinfo {author} {\bibfnamefont {J.~W.}\ \bibnamefont {Armstrong}},\ }\href
  {https://doi.org/10.1103/PhysRevD.66.122002} {\bibfield  {journal} {\bibinfo
  {journal} {Phys. Rev. D}\ }\textbf {\bibinfo {volume} {66}},\ \bibinfo
  {pages} {122002} (\bibinfo {year} {2002})},\ \Eprint
  {https://arxiv.org/abs/gr-qc/0209039} {arXiv:gr-qc/0209039} \BibitemShut
  {NoStop}%
\bibitem [{\citenamefont {Muratore}\ \emph {et~al.}(2022)\citenamefont
  {Muratore}, \citenamefont {Vetrugno}, \citenamefont {Vitale},\ and\
  \citenamefont {Hartwig}}]{Muratore:2021uqj}%
  \BibitemOpen
  \bibfield  {author} {\bibinfo {author} {\bibfnamefont {M.}~\bibnamefont
  {Muratore}}, \bibinfo {author} {\bibfnamefont {D.}~\bibnamefont {Vetrugno}},
  \bibinfo {author} {\bibfnamefont {S.}~\bibnamefont {Vitale}},\ and\ \bibinfo
  {author} {\bibfnamefont {O.}~\bibnamefont {Hartwig}},\ }\href
  {https://doi.org/10.1103/PhysRevD.105.023009} {\bibfield  {journal} {\bibinfo
   {journal} {Phys. Rev. D}\ }\textbf {\bibinfo {volume} {105}},\ \bibinfo
  {pages} {023009} (\bibinfo {year} {2022})},\ \Eprint
  {https://arxiv.org/abs/2108.02738} {arXiv:2108.02738 [gr-qc]} \BibitemShut
  {NoStop}%
\bibitem [{\citenamefont {Hartwig}\ and\ \citenamefont
  {Muratore}(2022)}]{Hartwig:2021mzw}%
  \BibitemOpen
  \bibfield  {author} {\bibinfo {author} {\bibfnamefont {O.}~\bibnamefont
  {Hartwig}}\ and\ \bibinfo {author} {\bibfnamefont {M.}~\bibnamefont
  {Muratore}},\ }\href {https://doi.org/10.1103/PhysRevD.105.062006} {\bibfield
   {journal} {\bibinfo  {journal} {Phys. Rev. D}\ }\textbf {\bibinfo {volume}
  {105}},\ \bibinfo {pages} {062006} (\bibinfo {year} {2022})},\ \Eprint
  {https://arxiv.org/abs/2111.00975} {arXiv:2111.00975 [gr-qc]} \BibitemShut
  {NoStop}%
\bibitem [{\citenamefont {Bayle}\ and\ \citenamefont
  {Hartwig}(2023)}]{BayleHartwig2023LISAInstrument}%
  \BibitemOpen
  \bibfield  {author} {\bibinfo {author} {\bibfnamefont {J.-B.}\ \bibnamefont
  {Bayle}}\ and\ \bibinfo {author} {\bibfnamefont {O.}~\bibnamefont
  {Hartwig}},\ }\href {https://doi.org/10.1103/PhysRevD.107.083019} {\bibfield
  {journal} {\bibinfo  {journal} {Phys. Rev. D}\ }\textbf {\bibinfo {volume}
  {107}},\ \bibinfo {pages} {083019} (\bibinfo {year} {2023})}\BibitemShut
  {NoStop}%
\end{thebibliography}%

\end{document}